\documentclass[twocolumn, a4paper, 11pt,book]{archive}
\usepackage{lineno,hyperref}
\modulolinenumbers[5]
\modulolinenumbers[1]
\usepackage{color}
\usepackage{multirow,bigdelim}
\usepackage{version}
\usepackage{graphicx}

\usepackage[]{natbib}

\usepackage{subcaption}
\usepackage{graphicx}
\usepackage{lscape}   
\usepackage{txfonts}
\usepackage{enumitem}   
\usepackage{xcolor}
\definecolor{darkorange}{rgb}{1.0, 0.55, 0.0}
\definecolor{britishracinggreen}{rgb}{0.0, 0.26, 0.15}
\definecolor{bondiblue}{rgb}{0.0, 0.58, 0.71}

\definecolor{klein}{rgb}{0.54, 0.17, 0.89}

\usepackage{soul}

\newcommand{\thisdisk}{Tau 042021}

\usepackage{tikz}
\usetikzlibrary{positioning}
\usepackage{pgfplots}
\usepackage{amsmath,amssymb,amsfonts}
\usepgfplotslibrary{fillbetween}
\begin{document} 

\twocolumn[{%
 \centering

  {\center \bf \Huge The edge-on disk Tau042021: icy grains at high altitudes and a wind containing astronomical PAHs}\\
\vspace*{0.25cm}

   {\Large E. Dartois \inst{1},          
   J. A. Noble
          \inst{2},
          M.~K. McClure
          \inst{3},
          J.~A. Sturm
\inst{3},
T.~L. Beck
\inst{4},
N. Arulanantham
\inst{5},
M.~N. Drozdovskaya
\inst{6},
C.~C. Espaillat
\inst{7,8},
D. Harsono
\inst{9},
M.-E. Palumbo
\inst{10},
Y.~J. Pendleton
\inst{11},
K.~M. Pontoppidan
\inst{5}
          }
          \vspace*{0.25cm}

   $^1$ Institut des Sciences Mol\'{e}culaires d’Orsay, CNRS, Univ. Paris-Saclay, 91405 Orsay, France\\
              \email{emmanuel.dartois@cnrs.fr}\\
   $^2$ Physique des Interactions Ioniques et Mol\'{e}culaires, CNRS, Aix Marseille Universit\'e, Marseille, France\\
   $^3$ Leiden Observatory, Leiden University, P.O. Box 9513, NL2300 RA Leiden, The Netherlands\\
   $^4$ Space Telescope Science Institute, 3700 San Martin Drive, Baltimore, MD 21218, USA\\
$^5$ Jet Propulsion Laboratory, California Institute of Technology, 4800 Oak Grove Drive, Pasadena, CA 91109, USA\\
$^6$ Physikalisch-Meteorologisches Observatorium Davos und Weltstrahlungszentrum (PMOD/WRC), Dorfstrasse 33, 7260, Davos Dorf, Switzerland\\
$^7$ Department of Astronomy, Boston University, 725 Commonwealth Avenue, Boston, MA 02215, USA\\
$^8$ Institute for Astrophysical Research, Boston University, 725 Commonwealth Avenue, Boston, MA 02215, USA\\
$^9$ Institute of Astronomy, Department of Physics, National Tsing Hua University, Hsinchu, Taiwan\\
$^10$ INAF—Osservatorio Astrofisico di Catania, via Santa Sofia 78, 95123 Catania, Italy\\
$^11$ Department of Physics, University of Central Florida, Orlando, FL 32816, USA\\

 \vspace*{0.5cm}

{\it \Large To appear in Astronomy \& Astrophysics}\\
 \vspace*{0.5cm}

}]

  \section*{Abstract}
    {Spectra of the nearly edge-on protoplanetary disks observed with the JWST have shown ice absorption bands of varying optical depths and peculiar profiles, challenging radiative transfer modelling and our understanding of dust and ice in disks.}
   {With the aim of constraining the underlying disk's structure and evolutionary state, we build models including dust grain size, shape, and composition to reproduce JWST IFU spectroscopy of a well-characterised, massive and large edge-on disk, Tau~042021. 
   Specifically we aim to match its spectral energy distribution, the spatial distribution of the dust and ice, the spectral characteristics of the dust continuum and ice bands profiles, as well as testing for the presence of astronomical PAH band carriers. }
{We explore radiative transfer models using different dust grain size distributions, including grains with effective radii $a_{\rm eff} = 0.005-3000\,\mu$m. Mass absorption and scattering coefficients for distributions of triaxial ellipsoidal grains are calculated using the discrete dipole approximation (DDA) for small size parameters ($2\pi a_{\rm eff}/\lambda < 10$), whereas the hollow sphere approximation is used for larger size parameters. We consider compositions including silicates, amorphous carbon, and mixtures of water, carbon dioxide, and carbon monoxide. The resulting orientation-averaged scattering matrices are input into RADMC-3D Monte Carlo radiative transfer models of Tau~042021 to simulate the spectral cubes observed with JWST-NIRSpec and MIRI. We compare the calculated optical depth distributions and profiles of the main ice bands to observations, including water at 3.05~$\mu$m, carbon monoxide at 4.67~$\mu$m, and carbon dioxide at 4.27 $\mu$m. We also compare these results to archival JWST-NIRCam and ALMA continuum images. We test three increasingly complex disk structures, starting from a standard model, first adding an extended atmosphere and then a disk wind containing astro-PAHs.}
{The observed near- to mid-infrared spectral energy distribution 
requires efficient scatterers, thus implying dust distributions that include grain sizes up to several tens of microns. 
The intensity distribution perpendicular to the disk exhibits emission profile wings extending into the upper disk atmosphere at altitudes exceeding the classical scale height expected in the isothermal hydrostatic limit. 
We produce ice absorption images that demonstrate the presence of icy dust grains up to altitudes high above the disk midplane, more than three hydrostatic equilibrium scale heights. 
We demonstrate the presence of a wind containing the carriers of astronomical PAH bands. The wind appears as an X-shaped emission at 3.3, 6.2, 7.7 and 11.3 microns, characteristic wavelengths associated with the infrared astronomical PAH bands. We associate the spatial distribution of this component with carriers of astronomical PAH bands that form a layer of emission at the interface with the H$_2$ wind.}
   {}




\section{Introduction}

Nearly edge-on protoplanetary disks offer the unique opportunity to trace the vertical distribution of dust and gas in a disk \citep[e.g.][]{Arulanantham2024, Duchene2024,Sturm2023b,Sturm2023a,Dutrey2017, Guilloteau2016}. However, at such high inclination, the extremely high opacity toward the central nascent star prohibits direct line of sight observation of ices in absorption in the visible to infrared range. The observed spectra rather result from a complex radiative transfer as light scattered on the disk surface probes different parts of the disk. Spectral features thus carry signatures of the disk's physical structure and of the different scattering properties of dust grains of various sizes, shapes, and compositions.
With JWST, the dust and ice composition of nearly edge-on disks can be spatially resolved to reveal the chemical stratification of the (icy) dust grains. Recently, Sturm et al. 2024 studied the radial and vertical distribution of major ice species toward the edge-on disk HH 48NE. 
That work demonstrated the retrieval of the spatial distribution of the ice is often non trivial due to the complex photon propagation through the disk.
The CO ice was observed at high altitudes similar to CO$_2$ and H$_2$O. Thorough radiative transfer models of icy dust grains are needed to infer both the ice and the dust grain distributions in the disk. 

The Tau~042021 disk (RA 04:20:21.4, Dec +28:13:49.2; 2MASS J04202144+2813491), is a large and nearly edge-on ($\sim$~88$^{\circ}$) disk, located in the nearby Taurus dense cloud \citep[$\sim$140 pc,][]{Galli2019, Kenyon1994, Roccatagliata2020}, first discovered in the visible \citep{Luhman2009, Stapelfeldt2014, Villenave2020}. Hubble Space Telescope Advanced Camera for Surveys (HST-ACS) observations further revealed the presence of a bipolar jet at 0.6 $\rm \mu m$ \citep{Duchene2014}. The central star is estimated to be an M1 star \citep{Luhman2009, Duchene2024}, with an effective temperature of $\sim$3700~K and a sub-solar bolometric luminosity. Along its major axis, the disk extends out to about 400--450 au in radius when observed in dust and gas in the visible and near-IR, or with CO gas in the millimetre, whereas the millimetric (larger grain) dust emission is slightly less extended \citep[][and references therein]{Duchene2024,Villenave2020}. 
It has been recently imaged in several infrared bands with the James Webb Space Telescope (JWST) by \cite{Duchene2024}, who analysed the images with refractory-only dust models, concluding that large $\sim$10\,$\mu$m dust grains must be present and vertically well-mixed with smaller grains. The large grains were found elevated to at least 60\,au above the midplane at the largest disk radii. This observation implies that turbulence plays a crucial role, being strong enough to lift such large particles all the way to the surface of the disk. 
The paper also discussed potential sources of emission observed high above the midplane in their images, including the presence of a photodissociative disk wind which could entrain grains as large as $\sim 1\,\mu$m, producing a `veil' of optical scattered light well above the disk midplane, and the possibility that PAHs/very small grains are entrained along with the gas and get stochastically heated once they reach high enough layers from which they are in direct sight of the central source.
In contrast, ALMA dust continuum images show a significant degree of settling for larger grains probed up to the millimeter size range \citep{Villenave2020, Villenave2023}.

In the first paper from the MIDAS JWST GO program (PID 1751), \citet{Arulanantham2024} used JWST MIRI spectroscopy to resolve the mid-IR spatial distribution of H$_2$, revealing an X-shaped disk wind extending to $\sim$200 au above the disk midplane, as well as a bipolar jet arising from [Ne II], [Ne III], [Ni II], [Fe II], [Ar II], and [S III] forbidden emission lines. Extended H$_2$O and CO gas emission lines were also detected, with excitation indicating they arise from the central regions close to the star, but are scattered by the large dust grains in the disk surface. Finally, extended astronomical PAH band emission at 11.3 $\rm \mu m$ co-spatial with the scattered light continuum was observed. 

In this article, we produce spatially resolved maps of ice absorption features in the \thisdisk disk and use radiative transfer models to simulate them. In section~\ref{Observations}, we present the NIRSpec spectra and ice maps. Section~\ref{Model} is devoted to the description of the physical model for the disk and describing how the icy grain properties are derived. A number of progressively complex physical models of spectra and ice maps are presented and compared to the data in Section~\ref{RT}. The final section presents the conclusions of the study.
%

\begin{figure*}[!ht]
\begin{center}
\begin{minipage}{0.6\columnwidth}
    \includegraphics[width=\columnwidth]{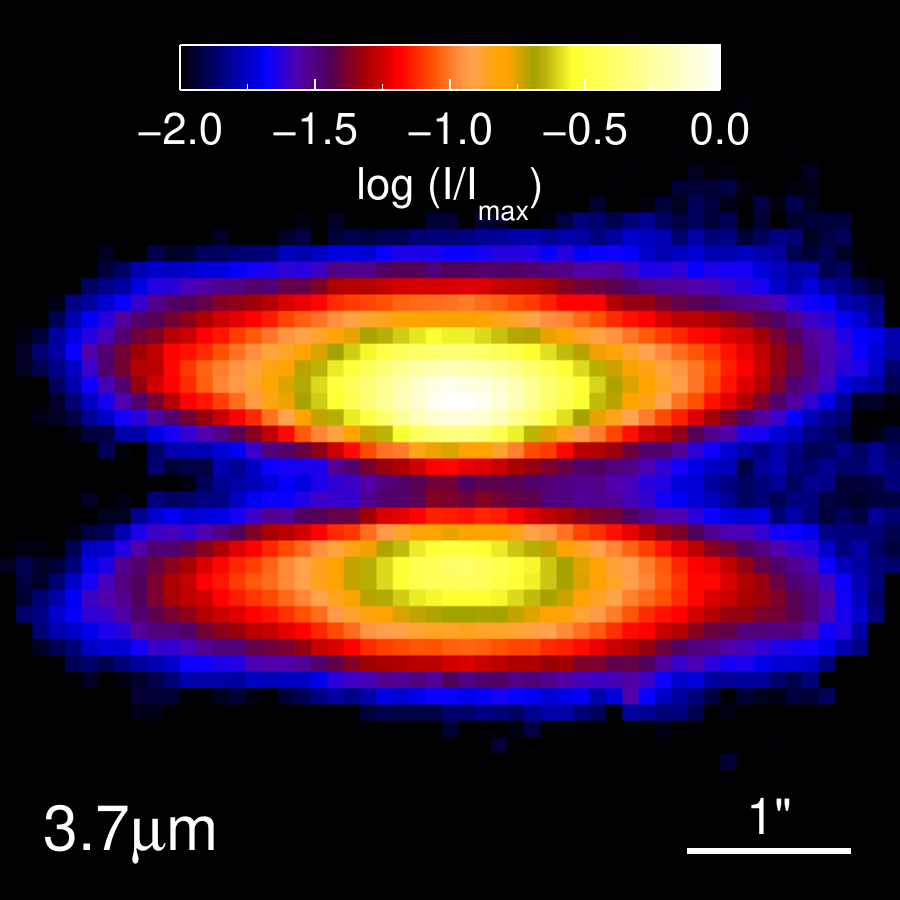}
\end{minipage}
\begin{minipage}{1.3\columnwidth}
\vspace*{0.5cm}
\includegraphics[width=\columnwidth,angle=0]{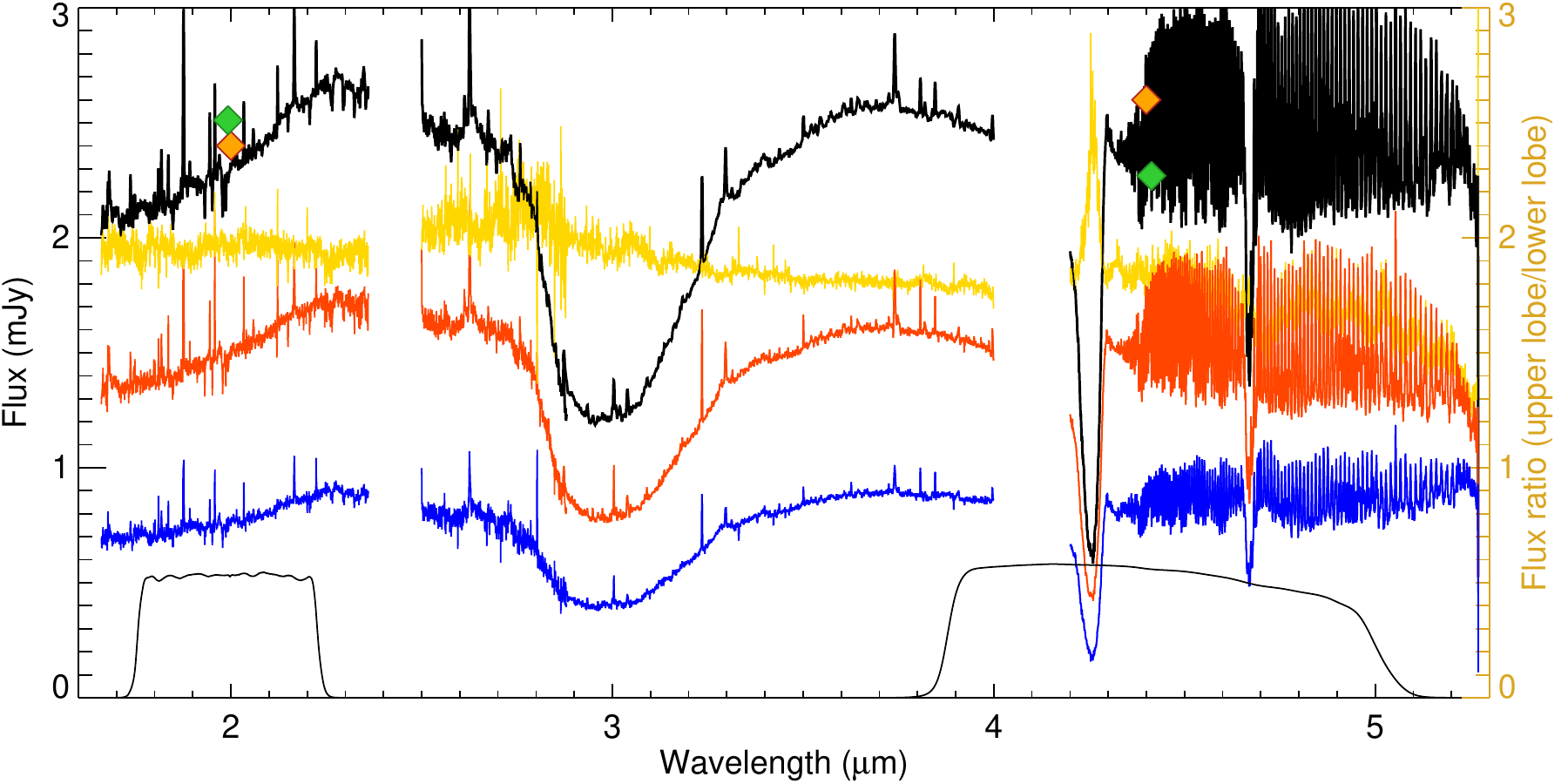}
\end{minipage}
\caption{
NIRSpec observations of \thisdisk. Left: An illustrative photometric image at 3.7~$\mu$m, showing the overall structure of the disk in continuum emission flux. Right: NIRSpec spectra of \thisdisk. The total flux integrated over both lobes is shown in black. The individual contributions of the upper emission lobe (red) and lower lobe (blue) are shown. The upper-to-lower lobe flux ratio is shown in gold. The ratio generally lies at around 2 across the continuum, whereas in some parts of the profile of the absorption bands, the lower lobe is more extinct than the upper.
Some gaps in the spectra are due to the corresponding ``gaps'' in the NIRSpec observations at some wavelengths. The photometry from NIRCam images presented in \cite{Duchene2024} are shown in orange (the corresponding filter shapes for the NIRCam filters F200W and F444W are shown below in black). The corresponding photometry obtained by integrating the NIRSpec spectra over the same filter profiles from our work are shown in green.}
\label{Figure_spectres_integres_NIRSpec}
\end{center}
\end{figure*}
%
\section{Observations}
\label{Observations}

The observations presented here are part of the JWST Cycle 1 GO program ``Mapping Inclined Disk Astrochemical Signatures (MIDAS)'' (\#1751, PI McClure). 

We observed Tau~042021 with NIRSpec on September 7th, 2023. The observations were taken with the G235H and G395H filters from 1.66-3.17 $\mu$m and 2.87-5.27 $\mu$m, respectively, at a spectral resolution of R$\sim$2700, using the IFU observing mode with a pixel scale of 0.1$\arcsec$~pixel$^{-1}$ and a FOV of 3.0$\arcsec x 3.0\arcsec$. To fully capture the radial and vertical extent of this disk, predicted by our pre-launch radiative transfer models, we used a 2x2 mosaic and a 4-point cycling pattern optimized for extended sources. Each cycled G395H exposure consisted of one integration of 25 groups with the NRSIRS2RAPID readout pattern for a total exposure time of 1518 s, while each cycled G235H exposure consisted of one integration of 5 groups with the NRSIRS2 readout pattern account, to account for data volume issues, for the same total exposure time of 1518 s. Since the regions surrounding Tau~042021 lacked bright stars in the field that could leak through the MSA failed open shutters and the intrinsic background emission is low at these wavelengths, we opted not to obtain either a leakcal frame or dedicated background for these NIRSpec observations.

For MIRI observations, we refer to details presented in \cite{Arulanantham2024} for which we recall here only a brief description. These observations were taken on February 13th, 2023. A 4-point dither pattern optimised for an extended source, with each dither exposure consisting of one integration of 69 groups, read using the FASTR1 readout pattern, for total on-source exposure times of 6594 s in each of the MIRI channels. A dedicated sky background was obtained with dithers of the same number of groups on a 2-point dither pattern. Observations used the four different IFU channels, covering four wavelength ranges: 4.9--7.65 $\mu$m (Channel 1), 7.51--11.7 $\mu$m (Channel 2), 11.55--17.98 $\mu$m (Channel 3), and 17.7--27.9 $\mu$m (Channel 4). The smallest FOV and highest spatial resolution are available at shorter wavelengths in Channel 1 ($3.2\arcsec \times 3.7\arcsec$, $0.196\arcsec$~pixel$^{-1}$). The FOV increases to $6.6\arcsec \times 7.7\arcsec$ in Channel 4, and the spatial resolution drops to $0.273\arcsec$~pixel$^{-1}$. For point sources, the average PSF FWHMs, denoted by theta in the subsequent equation, increase with wavelength, such that $\theta = 0.033 \left(\lambda \, \mu \rm{m} \right) + 0.106\arcsec$ \citep{Law2023, Argyriou2023}.

To supplement the MIDAS data and better constrain the physical model of Tau 042021, we gather archival multi-wavelength images and spectra. The disk Spectral Energy Distribution (SED) and HST/JWST/ALMA images were already modelled in \citet{Duchene2024} using dust grains based on photometry and broadband images. We extend the range of the NIRSpec and MIRI data by adopting the photometric points from Table~4 in \citet{Duchene2024}. In the visible to the near-infrared (NIR) wavelength range, it is known that the scattered light from the disk shows significant variability over a period of a few days \citep{Luhman2009}. This is seen when comparing photometric points recorded at different periods to the JWST photometry. Some degree of variability in the flux is thus expected in the NIR if the data are not recorded at the same epoch. 

\subsection{Observational extraction and analysis}

NIRSpec datacubes provided by the JWST pipeline (1188 pmap version) were processed with in-house post-pipeline treatments, including routines to flag glitches by comparing sudden localised ``hot'' pixels between adjacent channels, and interpolating the fluxes at the flagged spaxel positions. We applied a synthetic photometry approach to check the cross-calibration between NIRCam broadband images and NIRSpec IFU spectra. The NIRSpec spectro-photometry were compared to NIRCam by integrating through the F200W and F444W NIRCam bandpasses \citep{Rieke2023}, giving rise to photometric fluxes which agree well with \cite{Duchene2024} photometric fluxes, as shown in Fig.\ref{Figure_spectres_integres_NIRSpec}.
We extracted MIRI one-dimensional spectra by integrating over each lobe of the disk, after subtracting a background flux, estimated in an apparent empty region of the MIRI field of view. The spectra were scaled by 25\% to the NIRSpec flux density in the overlapping region around 5 $\mu$m to account for potential differences in extraction aperture and compensate for intrinsic source variability.
We construct optical depth images of the main ice bands by averaging the flux density of channels over 0.01~$\mu$m-wide bins at the wavelength center of each ice band. The optical depth of each spectral feature was derived by taking the ratio of the image at the ice band and an image at a continuum wavelength, chosen to be free of absorption. By taking an adjacent continuum point at a wavelength as close as possible to the ice band extinction, we take into account, as much as possible, the effects of propagation (and thus scattering) for similar wavelengths.
This operation proceeds only for spaxels with intensities above a given threshold above the noise, defining the accessible dynamic range presented in the images. In particular, this means we can not access the ice optical depth along the dark lane separating the two lobes. An optical depth map is derived as $\rm \tau(x,y) = -\ln(F_{\rm ice}(x,y)/F_{\rm cont}(x,y))$ to show ice spatial variations in ice distributions. 
If some photons are leaking from ice-free regions and contribute to the spectrum it adds an offset to the spectrum and implies the true optical depths might be higher. The important point is that these observed optical depth maps, as they will be compared to the models output ones constructed in the same way, allow a direct comparison between models and observations.

\subsection{Images and spectra}

Some NIRCam and MIRI images of Tau~042021 were shown in \cite{Duchene2024} and the MIRI spectrum has been shown in \cite{Arulanantham2024}, where the gas phase lines were analysed. 
The opaque disk midplane divides the disk emission into two lobes with an approximate flux ratio of two between the upper lobe and the lower one, as expected for such a highly inclined disk ($\sim$88$^{\circ}$). 
This is observed again in the left panel of Fig.~\ref{Figure_spectres_integres_NIRSpec}, where we present an image in the continuum at 3.7 $\mu$m derived from the NIRSpec datacube by averaging fluxes in the cube channels over a 0.02 $\mu$m span around the central wavelength.
The image is shown on a log scale, with the flux I normalised to the maximum ($\rm I_{max}$) in the map. The scale is shown in the colorbar above and we can note that it covers almost two orders of magnitude.

The NIRSpec spectrum is presented in the right panel of Fig.~\ref{Figure_spectres_integres_NIRSpec}. 
The emission from the two lobes is seen to maintain a relatively constant relative flux ratio of $\sim$2 across the spectrum. Strong ice absorption bands are evident in the spectrum of the disk, with the features dominated by the presence of H$_2$O, CO$_2$ and CO ices at 3.1, 4.27 and 4.67~$\mu$m, respectively. 
These are the major ice species expected to be present in ice mantles, and observed in other disks \citep[e.g.,][]{Sturm2023b}.
Both lobe spectra and the total flux spectrum display similar overall features, and particularly have similar ice band profiles.

In the spectral region that overlaps with the observations of this disk by \citet{Pascucci2024}, the fluxes we derive are in excellent agreement, and within potential source flux variations. As stated above, this disk scattering showed significant variability over a period of five days \citep[see Fig.10 in][]{Luhman2009} and between different epochs in 2MASS \citep{Cutri2003} 1.58 mJy at 2.2~$\mu$m, Spitzer, \citep{Rebull2010} 2.1 mJy at 3.6~$\mu$m, thus up to 60~\% variation, a variability much higher than potential JWST photometric calibration issues.
In the MIRI spectrum \citep[][and Fig.~\ref{Figure_gas_contributions}]{Arulanantham2024}, the main ice absorption feature observed is the bending mode absorption of CO$_2$ at about 15.2~$\mu$m.

In addition to the ice bands, emission lines are observed superimposed on the continuum in both the MIRI spectrum \citep[][and Fig.~\ref{Figure_gas_contributions}]{Arulanantham2024} and the NIRSpec spectrum in Fig.~\ref{Figure_spectres_integres_NIRSpec}. Between about 4.3 and 5.3~$\mu$m, on top of the continuum, arise intense, high temperature gas-phase CO emission lines coming from the inner disk regions. In the MIRI range, astronomical PAHs emission shapes the spectrum in the 6 to 13~$\rm\mu m$ range.
In this article, we will refer to these observed emission features as the astronomical PAH bands. However, in the literature, these are often referred to as Aromatic Infrared Bands (AIBs) since the exact carriers are not identified (i.e. they have not been attributed to specific polycyclic aromatic hydrocarbons). Some of the bands can definitively be attributed to the vibrational modes of aromatic species, and these features can also be referred to in the literature as `astro-PAHs', or even shortened to `PAH emission features'. In this article, to distinguish the observed emission features from the carriers of the bands, we will refer to astronomical PAH bands for the former and 
astronomical PAHs for the latter. 
When referring to model components used to reproduce both the emission and its associated carriers, we will use the term astro-PAHs.
\subsection{Optical depth imaging for H$_2$O, CO$_2$ and CO ices}
%
\begin{figure*}[h]
\begin{center}
\includegraphics[width=0.66\columnwidth,angle=0]{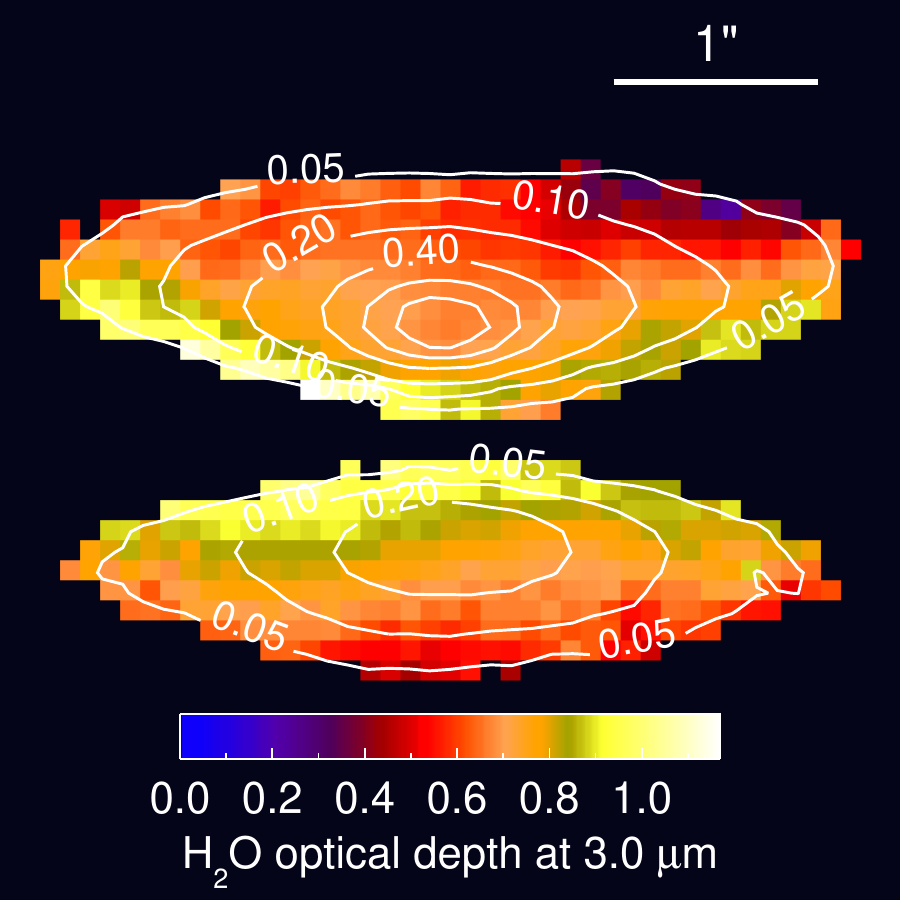}
\includegraphics[width=0.66\columnwidth,angle=0]{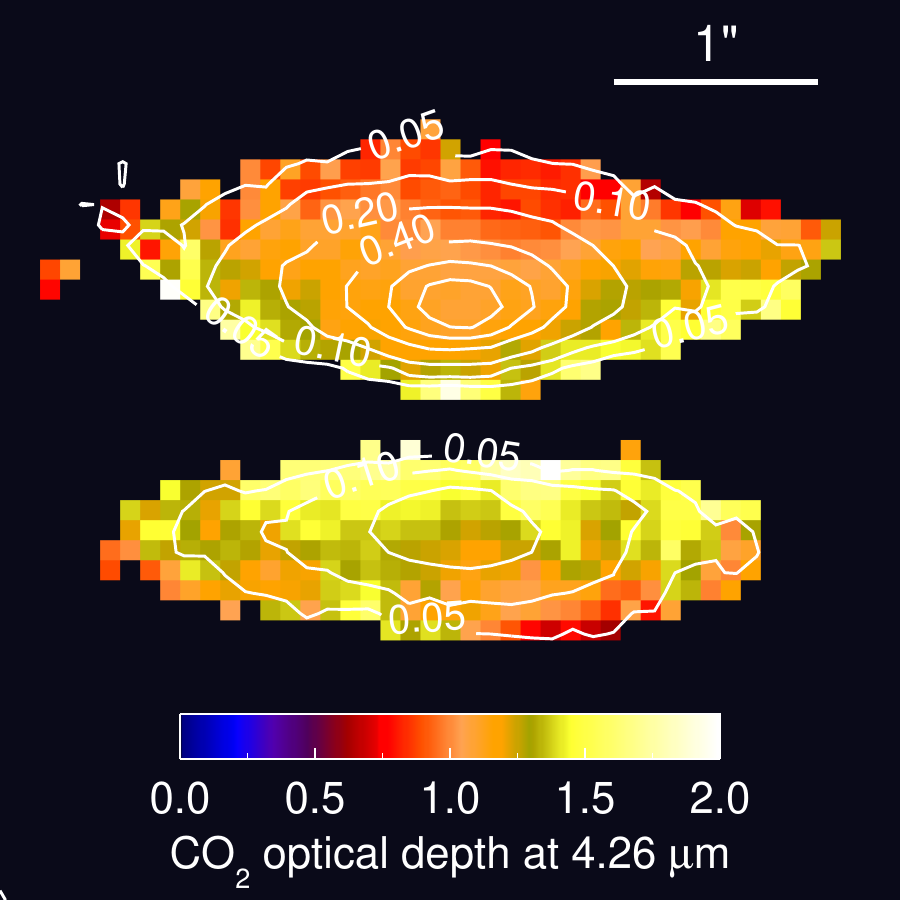}
\includegraphics[width=0.66\columnwidth,angle=0]{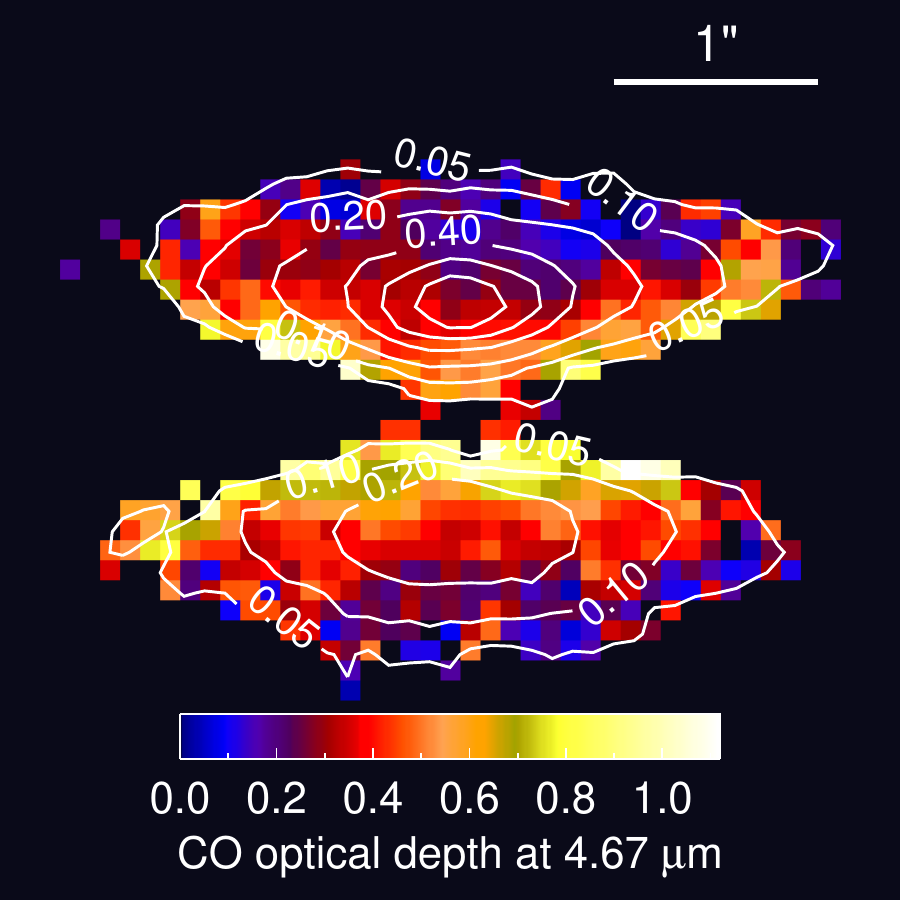}
\caption{Observed H$_2$O, CO$_2$ and CO ice optical depth mapped images of the disk at 3.0, 4.26, 4.67~$\mu$m derived from the NIRSpec image at those wavelengths calculated relative to the reference continuum intensities taken at 3.7, 4.32, 4.62~$\mu$m, respectively. Over-plotted white contours show the flux intensity at the band centres, normalised to their maximum in the map. The last contour is at 5\% from the maximum.}
\label{Figure_optical_depth_ices_obs}
\end{center}
\end{figure*}
%

Observed NIRSpec H$_2$O, CO$_2$ and CO optical depth mapped images of the disk at 3.0, 4.26, 4.67~$\mu$m calculated against the reference continuum intensities taken at 3.7, 4.32, 4.64~$\mu$m, respectively are shown in Fig.~\ref{Figure_optical_depth_ices_obs}. 
Over-plotted white contours show the relative intensities at the band centres, i.e., in the ice absorption, normalised to their maximum in the map. This shows the spatial extent of the dynamic range of the flux over which a meaningful optical depth could be derived.
A gradient in the optical depth is seen in every ice band, but the optical depths for H$_2$O, CO$_2$ ices do not go to zero up to the sensitivity limit of JWST, and are still present very high in the disk atmosphere.

The first striking aspect is that the ice band optical depth images seem to extend far from the disk midplane and are comparable, in terms of spatial profile, to the adjacent continuum. Note that the vertical extension of the images presented here is limited by the achieved JWST signal-to-noise, and thus, even if a significant decrease is observed in the optical depth, ices might be present at even greater distances from the midplane. Indeed, if the CO optical depth asymptotically goes to zero in the outermost parts of the atmosphere, for H$_2$O and CO$_2$ ices the optical depth are still significant when the disk intensity goes to the limit of detection by JWST, and they thus could be extending even slightly higher in the atmosphere than what we are able to measure here.
The lower lobe absorbs slightly more due to the inclination, although we note that optical imaging shows that the brightness asymmetry between lobes is variable \citep{Luhman2009}.  
\subsection{Spectral signatures of scattering in ice absorption profiles}
\begin{figure*}[h]
\begin{center}
\includegraphics[width=2\columnwidth,angle=0]{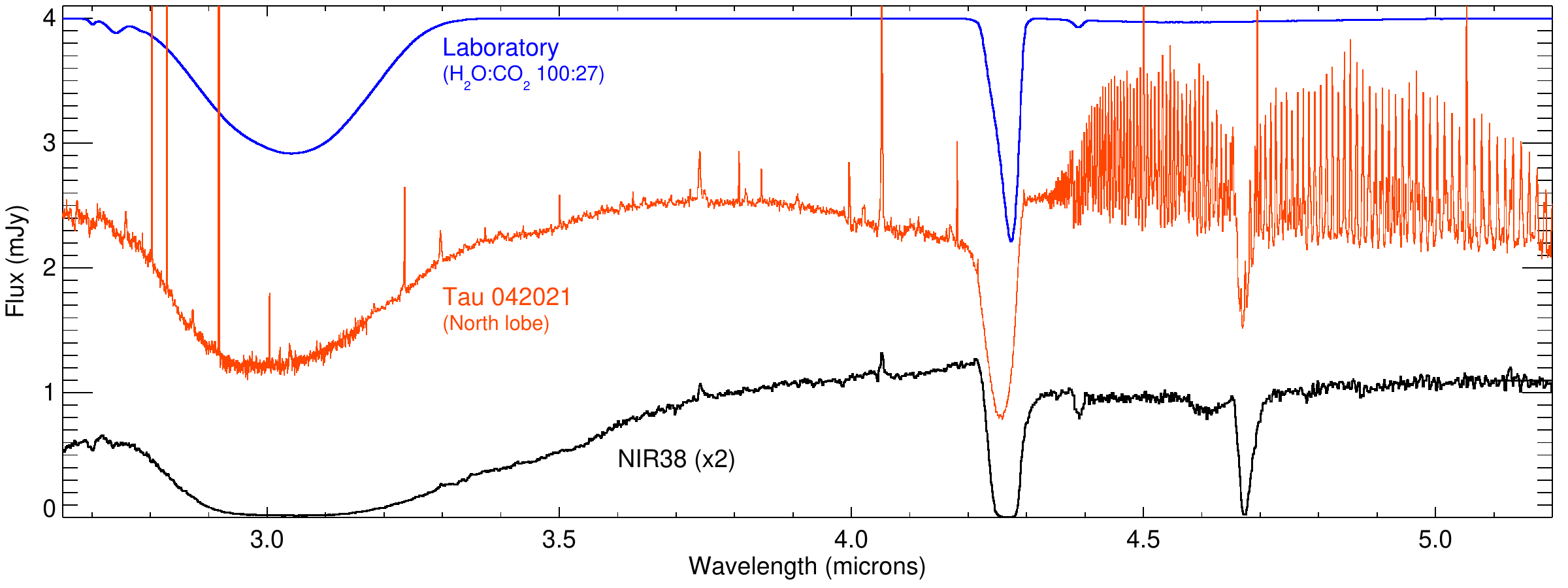}
\caption{Band profiles comparison. The integrated NIRSpec spectrum from the North lobe of \thisdisk, in a sub-region unaffected by NIRSpec IFU masking to fully dample the profile, is shown in red (multiplied by 1.2).
It is compared to a typical dense cloud spectrum (NIRCam + NIRSpec IFU, black, flux (multiplied by 2) observed at high extinction in Chamaeleon I \citep{McClure2023}. 
Note the inversion in the shape for the CO$_2$ ice band profile around 4.27~$\mu$m betraying differences in radiative transfer properties. A laboratory transmittance spectrum (blue) of a H$_2$O:CO$_2$ ice film mixture used in the optical constants determination in \cite{Dartois2024} has been converted to flux scale for comparison. Spectra have been scaled to have similar optical depths in the ice absorption features in order to compare their profiles.}
\label{Figure_profils}
\end{center}
\end{figure*}

The ice band profiles are unusual when compared to those generally observed in dense clouds or protostars with the JWST \cite[e.g.,][]{Yang2022,McClure2023,Federman2024, LeGouellec2024}, in particular for the CO$_2$ ice bands. A spectrum with complete wavelength coverage from the north emission lobe of \thisdisk in the NIR (i.e., excluding spaxels masked in the IFU at some wavelengths), scaled to the north lobe total flux at 3.7$\mu$m, is shown in Fig.~\ref{Figure_profils}. It is compared to a dense cloud spectrum from Chamaeleon I, taken from \citet{McClure2023}. 
The CO$_2$ antisymmetric stretching mode around 4.26~$\rm\mu m$ in this almost edge-on disk is distorted, with a long extinction wing on the blue side of the profile and a sharp increase on the red side, just before the comb of gaseous CO emission lines. 
This is reminiscent of the profile inversion observed in recent NIRSpec observations of HH~48~NE \citep{Sturm2023c} and several young objects with edge-on disks \citep[FS Tau B, HH 30, IRAS 04302, including \thisdisk~analysed here,][]{Pascucci2024}.
In contrast, the profile observed in the highly extinct source towards a dense cloud shows the opposite trend.
Such characteristic profile modifications can be attributed to radiative transfer effects, including scattering with respect to dense clouds. In the case of protoplanetary disks, as shown in, e.g., Fig.~14 of \citet{Dartois2022}, the emerging ice profiles will appear  more affected than for spherical envelopes or dense clouds. These profiles can even be reversed with respect to dense cloud ones, with a pronounced blue-shifted wing absorption, and especially for high inclination accompanied with high optical depths. Such profiles provide an important constraint on dust grains present in disks.
%
%
%
\section{Modelling and discussion}
\label{Model}

The observational data can be further interpreted with a series of models.  We focus particularly on the description of the dust grains, including the presence of ice, accounting for the observed ice features, before discussing the implications of adding two main additional structural components -- an extended dust grain atmosphere and a disk wind -- to the standard disk structure.

\subsection{Radiative transfer}
\label{RT}

To make relevant comparisons with the multiwavelength SED, JWST spectra and images,
we perform radiative transfer calculations to model the expected spectra and images for \thisdisk. The calculations are performed using the Monte Carlo RADMC-3D\footnote{https://www.ita.uni-heidelberg.de/$\sim$dullemond/software/radmc-3d/} software \citep{Dullemond2012}, utilizing full anisotropic scattering for the dust radiative transfer. 
Details on the grain models used, both in terms of size distributions and physico-chemical composition, are given in the following sections.
A spherical grid is used, and, taking advantage of the azimuthal symmetry, it is reduced to a radius and polar angle grid. The grid contains 256 radial points, spaced logarithmically in the radius from $\rm r_{in}$ to $\rm r_{out}$, and sampled even more finely near the inner disk edge. The polar angle is sliced into 512 angles covering a total opening angle of $\rm 3\pi /4$, centered on the disk midplane. 
After solving the radiative transfer, spectral cubes are produced at specific wavelengths with a FOV of $1000\times1000$~au, sampled with 5~au spaxels ($200\times200$ pixels), adopting a pre-defined disk inclination.
\subsection{Grain models}
The approach adopted here for modelling observed spectra of such icy grains was detailed in \cite{Dartois2022, Dartois2024}. For a comprehensive understanding of the concept development, we refer interested readers to these references. Here, we provide an overview of the various stages involved, as well as the addition made to extend the size range of the dust size distribution to millimetre-sized grains. We define the grain shape and size distributions, the composition of the refractory phase within those grains, identify the composition of the ice phase, and structure the arrangement of refractory and ice components within the final grain structure.

Each mixed ice-refractory grain is modelled using the Discrete Dipole Approximation (DDA) programme DDSCAT \citep[version 7.3,][]{Draine2013}. These grains are represented by dipoles with optical properties corresponding to both refractory and ice components, arranged stochastically to simulate a compact aggregate of smaller dust grains. 
This choice is driven by the expected stochastic sticking of dust grains from an initial
smaller sized distribution, each coated with an ice mantle, representing the continuous process, from dense cloud to the disk phase, of ice mantle production/accretion onto grains, acting concomitantly with collisions, leading to the aggregation into bigger grains \citep[e.g.][]{Marchand2023, Lebreuilly2023, Silsbee2020,  Paruta2016, Ormel2014}. Brief discussions on this choice can be found in \citet{Dartois2006, Dartois2022}.
Laboratory-derived optical constants of ices and refractory materials are used as described in \cite{Dartois2022, Dartois2024}. From the available laboratory ice mixtures, we can determine the optical constants. The ice mixture composition which proves to be closest to the main ice absorption features observed, i.e. the most adequate for modelling \thisdisk, corresponds to a binary H$_2$O:CO$_2$ ice combined with a pure CO ice (previously used in \cite{Dartois2024}). This choice has the additional advantage of showing how the radiative transfer affects differently the spectra in an edge-on disk when compared to a dense cloud line of sight, the latter discussed in \cite{Dartois2024}, starting from the same set of optical constants.
As dust grains exhibit non-spherical shapes, these add to the grain size in terms of contributing to scattering phenomena. 
To take this into account, we adopt a distribution of ellipsoids with many shapes represented by a weighted distribution of ellipsoid shapes exploring all the combinations of integers for the axis ratios within a factor of five (resulting in a total of 15 combinations). 
A discussion of this choice can be found in \S2.2 in \citet{Dartois2022}. In summary, such a distribution of ellipsoidal shapes explores a diversity of dust grains and, with such a weighting scheme, extreme shapes that have a non-physical probability of being present in space, such as when the ellipsoid tends to be like an infinite rod or to be planar, have a lower contribution to the distribution.
The orientation-averaged cross-section is obtained by averaging over sixteen orientation angles relative to the incoming plane wave for each ellipsoid. 
The DDA calculation is not suitable for calculating scattering and absorption cross sections for very large values of the dimensionless size parameter $\rm x = 2\pi a_{\text{eff}}/\lambda$, where  $a_{\text{eff}}$ is the effective radius of the grain and $\lambda$ the wavelength under consideration. At large size parameters, the absorption and scattering cross sections asymptotically converge towards their geometric regime.
For size parameters above ten, we use the cost effective approximation given by a distribution of hollow spheres \citep[$dhs$,][]{Min2005}, as implemented in the optool package \citep{Dominik2021}, to calculate the scattering and absorption cross sections for smaller wavelengths ($\lambda < 2\pi a_{\text{eff}}/10$). To ensure continuity between the calculations at their respective boundaries, the $dhs$ parameter (corresponding to $\rm f_{max}$, i.e. defining spheres with inner voids with volume fractions between 0 and $\rm f_{max}$) is adjusted at the wavelength $\lambda = 2\pi a_{\text{eff}}/10$ to match to, and thus provide continuity in the extinction cross sections calculated with, DDA.

The reference grain size distribution originates from the classical interstellar medium Mathis Rumpl \& Nordsieck (MRN) distribution \citep{MRN1977}, with grain radii ranging from $a_{\text{eff}}$(min) of 5$\times$10$^{-3}$ $\rm \mu m$ to $a_{\text{eff}}$(max) of 0.25 $\rm \mu m$, hereafter denoted as amin and amax, respectively, for simplicity. The number density of grains of size $a$, between these limits, follows a power law. For each dust size distribution, grain growth is simulated by redistributing the total dust mass into a new size distribution characterised by a given amin and amax.
Conserving the total dust mass thus redefines the slope of the power law in the distribution.
This approach for grain growth parametrization is consistent with numerous models' outcomes of grain growth up to the protostellar phase \citep{Marchand2022, Silsbee2020, Lebreuilly2019, Paruta2016, Ormel2014, Weingartner2001}.

\subsection{Dust grain components}

The standard dust model used in our disk models has three main components corresponding to distinct physical regimes in the disk. The dust grain existence limit is first defined by the sublimation temperature of silicates, which is set here to 1300~K, corresponding to their typical sublimation temperature in disks \citep[e.g.,][]{Kama2009}. At lower temperatures, the three components are as follows:\\
\begin{enumerate}[label=(\roman*)]
    \item The first grain composition, modelled as bare grains of a mix of silicates and amorphous carbon, is set by where the grains are heated and/or exposed to UV. Bare grains are expected for temperatures above the ice sublimation limit, set at 100~K, and below the refractory (i.e., silicates set at 1300~K) sublimation limit. In addition, at temperatures below 100~K, if the UV flux is high enough, photodesorption hampers the in-situ formation/re-accretion of a significant amount of ice . UV can come either from the central object or the exterior, so below a A$_{\text{V}}^{\text{th}}$, applied both radially and vertically bare grains are assumed. In practice, the ice sublimation is often the dominant parameter defining the radial limit of this first grain composition. We will return to the external UV field, i.e. vertical threshold, in the discussion.
   \item Another grain composition is defined below a visual extinction threshold, and where $T<100$~K, where grains contain a combination of the previously mentioned refractory components and ices with a composition as defined above.
   \item In the adopted dust grain size distributions, the largest grains start to decouple from the gas and progressively settle to the midplane of the disk \citep{Villenave2020}. To account for this, we thus implement an icy dust composition with a larger grain size distribution, which has the same chemical composition as (ii), but located at a smaller scale height within the disk. The icy grain distribution is thus divided into two ranges. The first size distribution, for grains fully dynamically coupled to the gas, extends from a$\rm_{min}$ to a$\rm_{settling}$, with a size distribution described by a power law adapted to preserve the total dust mass for an equivalent MRN distribution. This distribution's boundaries are adopted to bare grains and the icy small- to medium-sized grains. Above this size limit are the settled grains, and this large grain distribution extends from a$\rm_{settling}$ to 3 mm in size, the largest size limit in our calculations. The same power law is used for this distribution. An abundance factor is applied to the settled grain distribution to allow for relative variation of large grains with respect to the small- to medium-sized distribution. 
\end{enumerate}
%
%
\begin{figure}[h]
\begin{center}
\includegraphics[width=\columnwidth,angle=0]{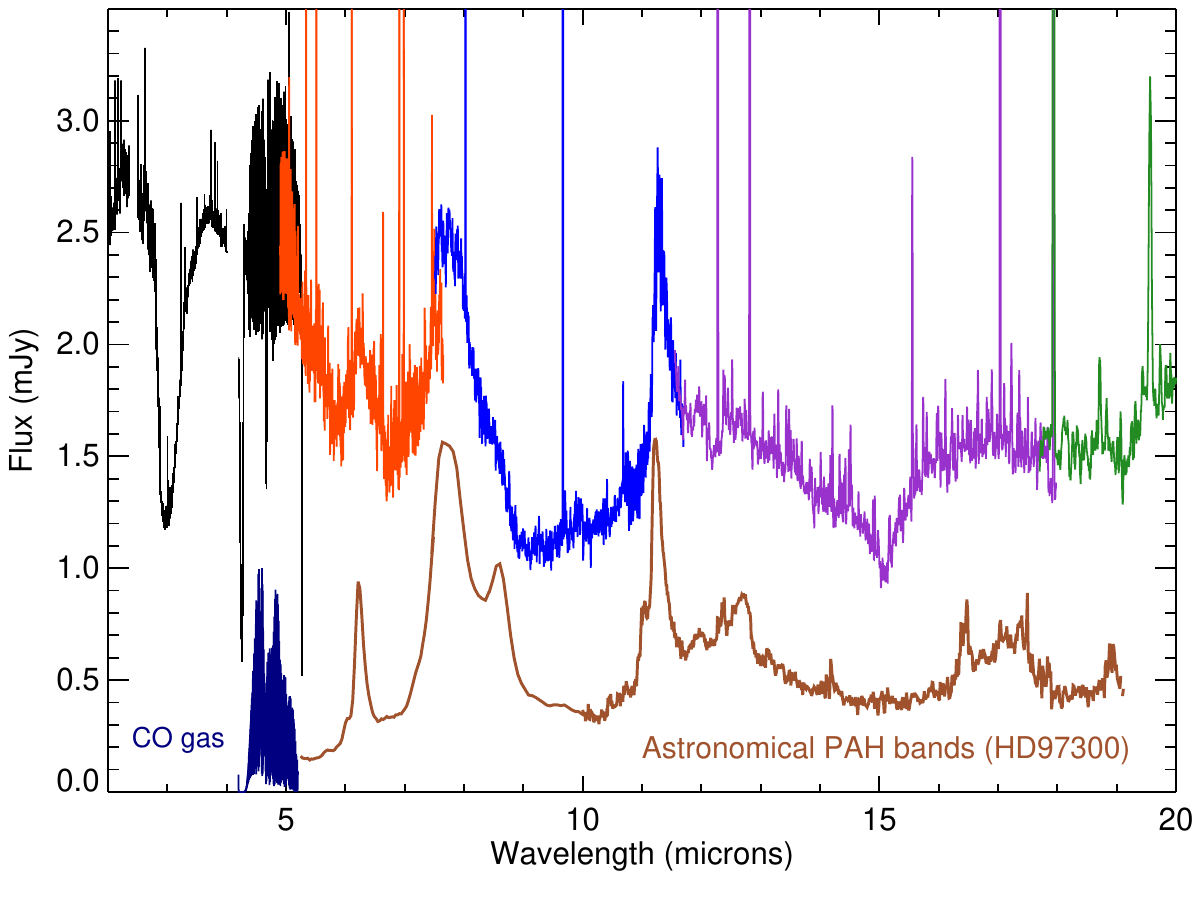}
\caption{Comparison of a combined NIRSpec (black) and MIRI (each channel is represented with a distinct colour: orange/blue/purple/green) spectrum of Tau~042021 with spectra representing gas-phase contributions from CO gas (dark blue) and astronomical PAHs (brown) to the spectrum. See text for details.  
}
\label{Figure_gas_contributions}
\end{center}
\end{figure}
%
%
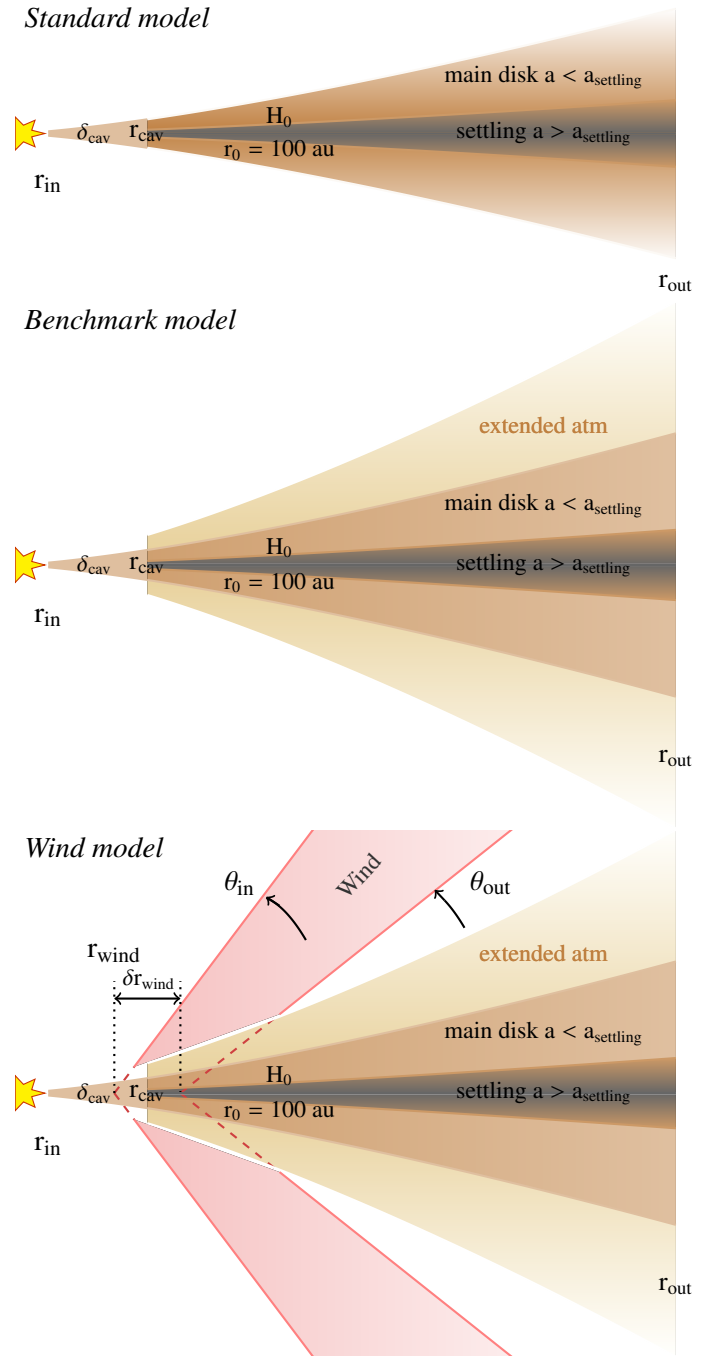
\begin{figure}[!ht]  
\begin{tikzpicture}[]
\begin{axis}[width  = \columnwidth*1.19,
height  = \columnwidth*0.7*0.95, xmin = 0, xmax = 105, ymin = -50, ymax = 50,samples=100,axis lines=none]
\addplot [red!80!green,fill=yellow,fill opacity=0.5,
    ] coordinates {
        (0,0)    (0,5)  (1.25,2.5)   (3.,4) (2.5,1) 
        (4.5,0) 
        (2.5,-1)    (3.,-4)  (1.25,-2.5)   (0,-5) (0,0) 
    };
\node[color=black, font=\large\bf\it, anchor=west] at (axis cs: 0,35) {Standard model};

\node[color=black, font=\normalsize] at (axis cs: 100,-45) {$\rm r_{out}$};

\addplot [domain=5:20, samples=100, name path=f, thick, color=brown!50] {0.075*x^1.35};
\addplot [domain=5:20, samples=100, name path=g, thick, color=brown!50] {-0.075*x^1.35};
\addplot[brown!50, opacity=0.5] fill between[of=f and g, soft clip={domain=5:20}];
\node[color=black, font=\large] at (axis cs: 5,-15) {$\rm r_{in}$};
%
\addplot [domain=20:100, samples=100, name path=f, thick, color=brown!95] {0.0*x^1.35};
\addplot [domain=20:100, samples=100, name path=g, thick, color=brown!5] {-0.075*x^1.35};
\addplot[shade, top color=brown!95, bottom color=brown!5, opacity=1.] fill between[of=f and g, soft clip={domain=20:100}];
\addplot [domain=20:100, samples=100, name path=f, thick, color=brown!95] {0.0*x^1.35};
\addplot [domain=20:100, samples=100, name path=g, thick, color=brown!5] {0.075*x^1.35};
\addplot[shade, bottom color=brown, top color=white, opacity=1] fill between[of=f and g, soft clip={domain=20:100}];
%
%
%
\node[color=black, font=\small] at (axis cs: 80,17) { main disk $\rm a<a_{settling}$};
 \addplot [domain=20:100, samples=100, name path=f, thick, color=black!60] {0.0*x^1.35};
\addplot [domain=20:100, samples=100, name path=g, thick, color=brown!80] {-0.02*x^1.35};
\addplot[shade, shading angle=90, anchor=west, top color=black!60, bottom color=brown!80, opacity=0.1] fill between[of=f and g, soft clip={domain=20:100}];
 \addplot [domain=20:100, samples=100, name path=f, thick, color=black!60] {0.0*x^1.35};
\addplot [domain=20:100, samples=100, name path=g, thick, color=brown!80] {0.02*x^1.35};
\addplot[shade, shading angle=90, anchor=west, bottom color=black!60, top color=brown!80, opacity=0.1] fill between[of=f and g, soft clip={domain=20:100}];
%
%
%
\node[color=black, font=\small] at (axis cs: 80,0) { settling $\rm a>a_{settling}$};

\node[color=black, font=\small] at (axis cs: 40,-5) {$\rm r_{0}=100$ au};
\node[color=black, font=\small] at (axis cs: 40,5) {$\rm H_{0}$};
\node[color=black, font=\normalsize] at (axis cs: 20,0) {$\rm r_{cav}$};
\node[color=black, font=\small] at (axis cs: 12,0) {$\rm \delta_{cav}$};
%

    \end{axis}
\end{tikzpicture}
%
%
%
\begin{tikzpicture}[]
\begin{axis}[width  = \columnwidth*1.19, height=\columnwidth*0.95, xmin = 0, xmax = 105, ymin = -75, ymax = 75,samples=100,axis lines=none]
\addplot [red!80!green,fill=yellow,fill opacity=0.5,
    ] coordinates {
        (0,0)    (0,5)  (1.25,2.5)   (3.,4) (2.5,1) 
        (4.5,0) 
        (2.5,-1)    (3.,-4)  (1.25,-2.5)   (0,-5) (0,0) 
    };
\node[color=black, font=\large\bf\it, anchor=west] at (axis cs: 0,70) {Benchmark model};

\node[color=black, font=\normalsize] at (axis cs: 100,-55) {$\rm r_{out}$};
\addplot [domain=20:100, samples=100, name path=f, thick, color=brown!0] {0.15*x^1.35};
\addplot [domain=20:100, samples=100, name path=g, thick, color=brown!0] {0*x^1.35};
\addplot[shade, bottom color=brown!75!yellow!50, top color=yellow!1, opacity=1.] fill between[of=f and g, soft clip={domain=20:100}];
\node[color=brown, font=\small] at (axis cs: 80,40) {extended atm};
\addplot [domain=20:100, samples=100, name path=f, thick, color=brown!0] {0*x^1.35};
\addplot [domain=20:100, samples=100, name path=g, thick, color=brown!0] {-0.15*x^1.35};
\addplot[shade, top color=brown!75!yellow!50, bottom color=yellow!1, opacity=1.] fill between[of=f and g, soft clip={domain=20:100}];
\node[color=brown, font=\small] at (axis cs: 80,40) {extended atm};

\addplot [domain=5:20, samples=100, name path=f, thick, color=brown!50] {0.075*x^1.35};
\addplot [domain=5:20, samples=100, name path=g, thick, color=brown!50] {-0.075*x^1.35};
\addplot[brown!50, opacity=0.5] fill between[of=f and g, soft clip={domain=5:20}];
\node[color=black, font=\large] at (axis cs: 5,-15) {$\rm r_{in}$};
%
\addplot [domain=20:100, samples=100, name path=f, thick, color=brown!50] {-0.075*x^1.35};
\addplot [domain=20:100, samples=100, name path=g, thick, color=brown!50] {0.075*x^1.35};
\addplot[shade, left color=brown!75, right color=brown!50, opacity=1.] fill between[of=f and g, soft clip={domain=20:100}];
\node[color=black, font=\small] at (axis cs: 80,17) { main disk $\rm a<a_{settling}$};
 \addplot [domain=20:100, samples=100, name path=f, thick, color=black!60] {0.0*x^1.35};
\addplot [domain=20:100, samples=100, name path=g, thick, color=brown!80] {-0.02*x^1.35};
\addplot[shade, top color=black!60, bottom color=brown!80, opacity=0.1] fill between[of=f and g, soft clip={domain=20:100}];
 \addplot [domain=20:100, samples=100, name path=f, thick, color=black!60] {0.0*x^1.35};
\addplot [domain=20:100, samples=100, name path=g, thick, color=brown!80] {0.02*x^1.35};
\addplot[shade, bottom color=black!60, top color=brown!80, opacity=0.1] fill between[of=f and g, soft clip={domain=20:100}];

\node[color=black, font=\small] at (axis cs: 80,0) { settling $\rm a>a_{settling}$};

\node[color=black, font=\small] at (axis cs: 40,-5) {$\rm r_{0}=100$ au};
\node[color=black, font=\small] at (axis cs: 40,5) {$\rm H_{0}$};
\node[color=black, font=\normalsize] at (axis cs: 20,0) {$\rm r_{cav}$};
\node[color=black, font=\small] at (axis cs: 12,0) {$\rm \delta_{cav}$};


    \end{axis}
\end{tikzpicture}
%
%
\begin{tikzpicture}[]
\begin{axis}[width  = \columnwidth*1.19, height=\columnwidth*0.95, xmin = 0, xmax = 105, ymin = -75, ymax = 75,samples=100,axis lines=none]
\addplot [red!80!green,fill=yellow,fill opacity=0.5,
    ] coordinates {
        (0,0)    (0,5)  (1.25,2.5)   (3.,4) (2.5,1) 
        (4.5,0) 
        (2.5,-1)    (3.,-4)  (1.25,-2.5)   (0,-5) (0,0) 
    };
\node[color=black, font=\large\bf\it, anchor=west] at (axis cs: 0,70) {Wind model};
\addplot [domain=18:100, samples=100, name path=f, thick, color=red!50] {2.5*(x-15)};
\addplot [domain=15:18, samples=100, name path=ff, thick, dashed, color=red!75!black!75, mark options={solid,draw=black}] {2.5*(x-15)};
\addplot [domain=40:100, samples=100, name path=g, thick, color=red!50] {1.5*(x-25)};
\addplot [domain=25:40, samples=100, name path=gg, thick, dashed, color=red!75!black!75, mark options={solid,draw=black}] {1.5*(x-25)};
\addplot[shade, left color=black!15!red!25, right color=white!50, opacity=1] fill between[of=f and g, soft clip={domain=10:100}];
\addplot [domain=18:100, samples=100, name path=f, thick, color=red!50] {-2.5*(x-15)};
\addplot [domain=15:18, samples=100, name path=ff, thick, dashed, color=red!75!black!75, mark options={solid,draw=black}] {-2.5*(x-15)};
\addplot [domain=40:100, samples=100, name path=g, thick, color=red!50] {-1.5*(x-25)};
\addplot [domain=25:40, samples=100, name path=gg, thick, dashed, color=red!75!black!75, mark options={solid,draw=black}] {-1.5*(x-25)};
\addplot[shade, left color=black!15!red!25, right color=white!50, opacity=1] fill between[of=f and g, soft clip={domain=10:100}];
\node[color=black, font=\large, rotate=0] at (axis cs: 10+5,25+15) {$\rm r_{wind}$};
\addplot[thick, dotted, mark options={solid,draw=black}] coordinates {(15,0.0) (15.,32)};
\addplot[thick, dotted, mark options={solid,draw=black}] coordinates {(25,0.0) (25.,32)};
\node[color=black, font=\small, rotate=0] at (axis cs: 28-7.75,25+7.5) {$\rm \delta r_{wind}$};
\addplot[<->, thick, mark options={solid,draw=black}] coordinates {(15,27.0) (25.,27)};
\node[color=black, font=\large, rotate=0] at (axis cs: 34,60) {$\rm \theta_{in}$};
\node[color=black, font=\large, rotate=0] at (axis cs: 72,60) {$\rm \theta_{out}$};
\node[color=black!80, font=\small, rotate=47] at (axis cs: 52,62) {Wind};

\node[color=black, font=\normalsize] at (axis cs: 100,-55) {$\rm r_{out}$};
\addplot [domain=20:100, samples=100, name path=f, thick, color=brown!0] {0.15*x^1.35};
\addplot [domain=20:100, samples=100, name path=g, thick, color=brown!0] {0*x^1.35};
\addplot[shade, bottom color=brown!75!yellow!50, top color=yellow!1, opacity=1.] fill between[of=f and g, soft clip={domain=20:100}];
\node[color=brown, font=\small] at (axis cs: 80,40) {extended atm};
\addplot [domain=20:100, samples=100, name path=f, thick, color=brown!0] {0*x^1.35};
\addplot [domain=20:100, samples=100, name path=g, thick, color=brown!0] {-0.15*x^1.35};
\addplot[shade, top color=brown!75!yellow!50, bottom color=yellow!1, opacity=1.] fill between[of=f and g, soft clip={domain=20:100}];
\node[color=brown, font=\small] at (axis cs: 80,40) {extended atm};

\addplot [domain=5:20, samples=100, name path=f, thick, color=brown!50] {0.075*x^1.35};
\addplot [domain=5:20, samples=100, name path=g, thick, color=brown!50] {-0.075*x^1.35};
\addplot[brown!50, opacity=0.5] fill between[of=f and g, soft clip={domain=5:20}];
\node[color=black, font=\large] at (axis cs: 5,-15) {$\rm r_{in}$};
%
\addplot [domain=20:100, samples=100, name path=f, thick, color=brown!50] {-0.075*x^1.35};
\addplot [domain=20:100, samples=100, name path=g, thick, color=brown!50] {0.075*x^1.35};
\addplot[shade, left color=brown!75, right color=brown!50, opacity=1.] fill between[of=f and g, soft clip={domain=20:100}];
\node[color=black, font=\small] at (axis cs: 80,17) { main disk $\rm a<a_{settling}$};
 \addplot [domain=20:100, samples=100, name path=f, thick, color=black!60] {0.0*x^1.35};
\addplot [domain=20:100, samples=100, name path=g, thick, color=brown!80] {-0.02*x^1.35};
\addplot[shade, top color=black!60, bottom color=brown!80, opacity=0.1] fill between[of=f and g, soft clip={domain=20:100}];
 \addplot [domain=20:100, samples=100, name path=f, thick, color=black!60] {0.0*x^1.35};
\addplot [domain=20:100, samples=100, name path=g, thick, color=brown!80] {0.02*x^1.35};
\addplot[shade, bottom color=black!60, top color=brown!80, opacity=0.1] fill between[of=f and g, soft clip={domain=20:100}];

\node[color=black, font=\small] at (axis cs: 80,0) { settling $\rm a>a_{settling}$};

\node[color=black, font=\small] at (axis cs: 40,-5) {$\rm r_{0}=100$ au};
\node[color=black, font=\small] at (axis cs: 40,5) {$\rm H_{0}$};
\node[color=black, font=\normalsize] at (axis cs: 20,0) {$\rm r_{cav}$};
\node[color=black, font=\small] at (axis cs: 12,0) {$\rm \delta_{cav}$};

%
\addplot [domain=65:68, samples=100, name path=f, thick, color=black] {60.-0.2*(60-x)^2};
\addplot[<-, thick, mark options={solid,draw=black}] coordinates {(63.5,58.0) (65.,55)};
\addplot [domain=38:44, samples=100, name path=f, thick, color=black] {56.5-0.2*(36-x)^2};
\addplot[<-, thick, mark options={solid,draw=black}] coordinates {(38,56.0) (40.,53.5)};


    \end{axis}
\end{tikzpicture}
\caption{Schematic edge-on views of the protoplanetary disk models used in this article. Upper panel: The `standard' model includes a main disk flaring component with grains of sizes smaller than a critical size a$\rm_{settling}$ and, possibly, settled dust grains for sizes above this size limit. The internal part of the disk (below r$_{cav}$) can harbour a partially filled cavity (where the density is multiplied by a $\delta_{cav}<1$ factor). See \S~\ref{Standard_disk_structure} for details of the other parameters used in the `standard model'.
Central panel: The extended atmosphere model, or `benchmark' model, detailed in  \S~\ref{benchmark_with_extended_atmosphere}, supersedes the `standard' model, while still including its main disk flaring component and settled dust grains. This model comprises an extended tenuous dust atmosphere lying above the classical purely hydrostatic model.
Lower panel: The `wind' model, detailed in  \S~\ref{wind_model_containing_astro_PAH}, includes all the components from the benchmark model with the addition of a wind whose main parametrisation is shown here. Angles are given from the midplane axis.}
\label{figures_schematiques}
\end{figure}
%
\subsection{Gas phase CO and astronomical PAH bands treatment}

Gas is included in the disk models as follows: a gas-to-dust ratio of one hundred is adopted, and the disk density is defined with respect to the gas distribution.
 In this section, we describe two components which present key spectral signatures in the observed data. A gas-phase CO component is particularly evident in the observed spectra, with CO lines spanning the $\sim4.3-5.3$~~$\mu$m range, covered by NIRSpec (Fig.~\ref{Figure_gas_contributions}).
We do not explicitly include the spectral signatures of this CO gas in the radiative transfer model.
Fig.~\ref{Figure_gas_contributions} shows a simulated spectrum of gas-phase CO from the inner region at $T\sim1500$~K. 
Such a gas-phase contribution is produced in regions ($r < 10$ au and/or high above the disk surface; \citealt{Salyk2011}), which can be, to first order, decoupled from the core of the main (icy) disk radiative transfer model. CO gas is out of the main scope of this article and will be analyzed in future studies.
The observed astronomical PAH bands emission spectrum from the HD97300 disk retrieved from the Spitzer archive (PID 2; Astronomical Observation Request 12697088, Low and High resolution), is also shown in Fig.~\ref{Figure_gas_contributions}. 
From the central positions and widths of the bands, it can be seen that the astronomical PAH bands from the HD97300 disk belong to a similar class of emission profiles as those in \thisdisk, thus allowing a first-order evaluation of their contribution to the SED. Astro-PAHs are not included in the first stages of the radiative transfer modelling, but will be discussed in a dedicated section below.

\subsection{Standard disk structure}
\label{Standard_disk_structure}

In order to analyse the observations, we use a series of disk models with increasing levels of structural complexity. 
We start with a classical, so-called `standard' disk model, as illustrated in the upper sketch of Fig.~\ref{figures_schematiques}.
The disk model is based on an axisymmetric flaring disk around a young star, described using classical parameters.
The surface density of the disk is parametrised with
\begin{equation}
\rm    
\Sigma(r) = \Sigma_0 
\left(\frac{r}{r_{\text{0}}}\right)^{-p} 
\exp\left(-\left(\frac{r}{r_t}\right)^{2-p}\right)
\end{equation}
where r is the radial distance to the star, $r_{\text{0}}$ is a reference radius and $\rm \Sigma_0$ the surface density normalisation at this reference radius. p is the power law exponent describing the radial variation, positive if the surface density decreases with radius. The density is tapered by an exponential edge with reference radius $r_{\text{t}}$, following the viscous disk model theoretical solution (Lynden-Bell \& Pringle 1974).

The vertical density distribution, under hydrostatic equilibrium, is given by
\begin{equation}
\rm        \rho(r, z) = 
\frac{\Sigma(r)}{H(r)\sqrt{2\pi}}
\exp\left(- z^2 / 2 H^2(r)\right)
\label{eqn_vertical_density_hydro}
\end{equation}
with the density at radial distance $\rm r$ and vertical distance $\rm z$ from the midplane,
and where $\rm H(r)$ is the vertical scale height under a vertical isothermal hypothesis, whose radial variation is given by
\begin{equation}
H(r) = H_0 \left(\frac{r}{r_0}\right)^{h}
\end{equation}
where $\rm H_0$ is the scale height at the reference radius $\rm r_0$. h is the power law exponent describing the scale height radial variation, $>1$ for flaring disks.
The midplane density radial variation evolves thus with a power law as $\rm s = -(p + h)$.\\

The disk extends from its inner radius $\rm r_{in}$, defined by the sublimation temperature of the refractory components to the outer radius $\rm r_{out}$ that we set to 450 au, based on previous JWST and ALMA imaging \citep[e.g.][]{Villenave2020,Arulanantham2024}, based upon a distance to the Taurus cloud of 140 pc \citep{Kenyon1994,Roccatagliata2020,Galli2019}.
Rather than a sharp sublimation front, many disks possess an inner, less dense cavity filled with lower density of material. We add to the model the possible presence of an inner cavity, which then extends from $\rm r_{in}$ to $\rm r_{cav}$, as discussed in, e.g., \cite{Sturm2023a}, and which is only partly filled with grains by applying a $\rm \delta_{cav}$ ($<<1$) factor to the prescribed densities for all the grain populations present at radii lower than $\rm r_{cav}$. 
To avoid overly sharp edges at the cavity boundary, the transition from the cavity edge to the $\rm \delta_{cav}$ factor inside the cavity is tapered over a few percent of the cavity radius.

Millimetre images have also shown that large grains are progressively settling towards the midplane. To take into account this observed settling, the dust grain distribution is divided into two grain populations, one for small to medium grains and another one for large grains, as explained above. The large grain distribution possesses its own (over)density factor with respect to the small grain distribution, and its own (lower) scale height $\rm H^L_0$.
Each of these size distributions is homogeneous, and is well-mixed within each zone.

\begin{figure*}[!ht]
\begin{center}
\includegraphics[width=2\columnwidth,angle=0]{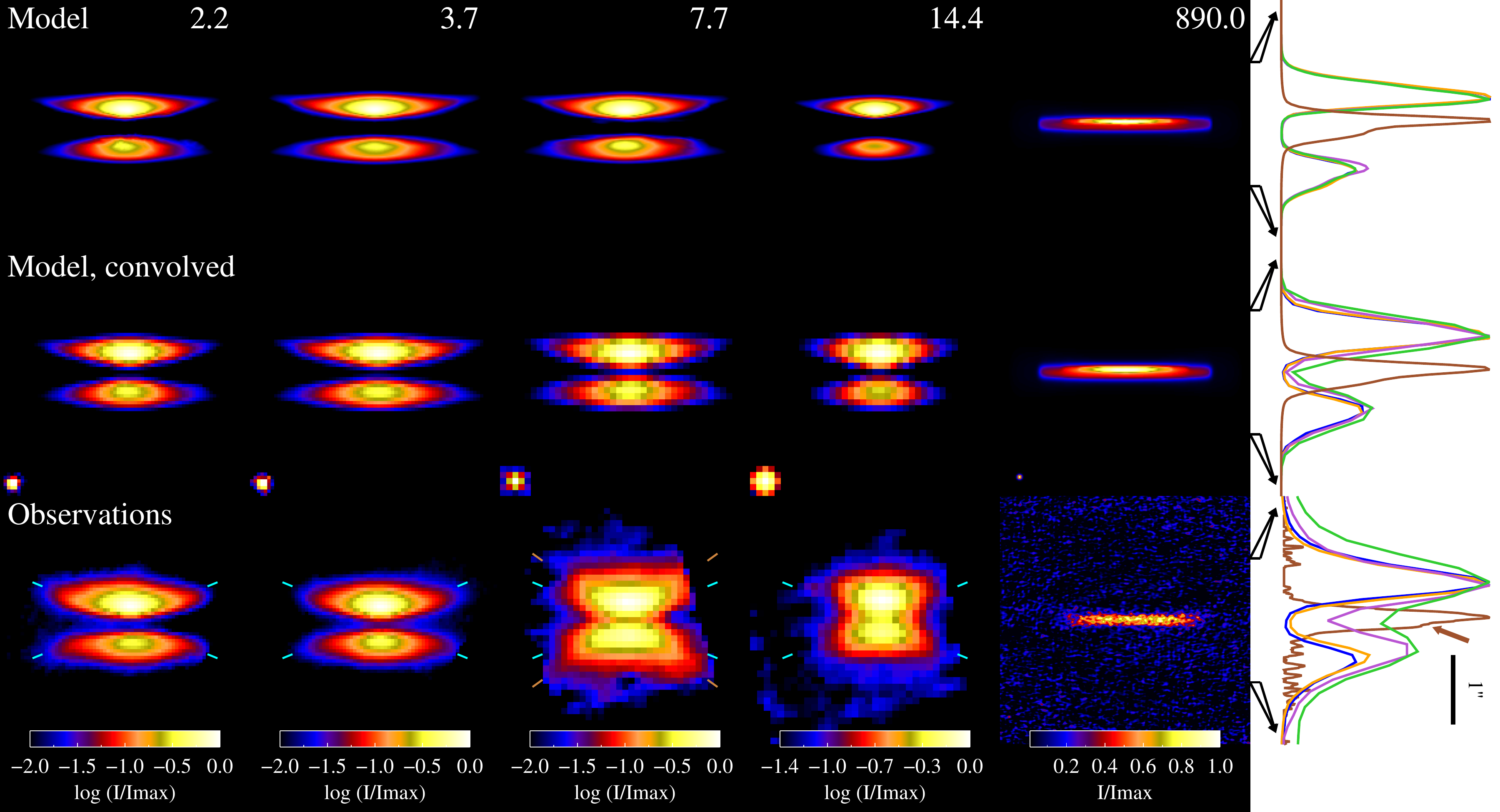}
\caption{Standard model. (upper) Model images at full resolution for selected wavelengths spanning the NIR to mm range. (middle) Model images once convolved with the JWST or ALMA PSF. (lower) JWST NIRSpec (2.2 and 3.7~$\mu$m), MIRI (7.7 and 14.4~$\mu$m), and ALMA Band 7 (890~$\mu$m) observations.
Blue ticks are guides to the small wings observed at about $22.5^{\circ}$, especially in the near infrared, whereas brown tickmarks indicate an angle of $36^{\circ}$ overplotted on the 7.7 $\mu$m image, corresponding to the angle for the X-shape discussed by \cite{Duchene2024}, and close to the H$_2$ wind (semi-opening) angle previously observed in the $35-38.5^{\circ}$ range by \cite{Arulanantham2024, Pascucci2024}.  
The colorbars indicate the intensity levels, normalised to the maximum intensity for each image, in log scale except for ALMA data presented in linear scale. The right panel are cuts along the vertical line through the centre of the disk observations and models at the corresponding wavelengths (2.2~$\mu$m - blue; 3.7~$\mu$m -orange; 7.7~$\mu$m - purple; 14.4~$\mu$m - green, and 890~$\mu$m - brown). The spatial scale is expanded (shown with the black arrows) by a factor of two  as compared to the images for a clearer view. The adopted blue-red-yellow color table was chosen to enhance the visibility of details in the structure. A version of this same figure with a uniform red temperature color table is presented in the appendix.}
\label{Figure_standard_images}
\end{center}
\end{figure*}

The results of a standard model -- with the main parameters close to those used in \cite{Duchene2024}, i.e., T$_{star}= 3700$~K, inclination of 87.5$^{\circ}$, $n_{\text{H}_{2}}(100~\text{au}) = 1\times10^9$~cm$^{-3}$, $\rm p = 0.7$, $\rm r_{in}$ = 0.07 au, $\rm r_{out} = 450 au$, $\rm H_{0}$ = $7.5$ au -- are presented in Fig.~\ref{Figure_standard_images} alongside observational data. The model includes, in addition, a lower density cavity $\rm r_{cav}$ = 40 au, $\rm\delta_{cav} = 0.05$.
The grain size distributions are set with a$\rm_{min}$ =1 $\rm\mu$m, a$\rm_{settled}$ = 40 $\rm\mu$m, a$\rm_{max}$ = 3000 $\rm\mu$m,  $\rm h = 1.2$. The settled large grain scale height in this model ($\rm H_{0}^{settled}$ = $\rm H_{0}/4$) is less pronounced than that used previously \citep[e.g.][]{Villenave2020}, due to the requirement for a slight asymmetry, which can be seen in the ALMA Band 7 image cut. The settled grain outer radius is also reduced as compared to the small dust and gaseous disk, here $0.7\times\rm r_{out}$, which is compatible with previous studies using a settled radius of about 300 au \citep{Villenave2020,Duchene2024}.

The first striking difference of such a ``standard'' model with the observations is the vertical extent of the disk, particularly evident at lower wavelengths. Even after the PSF convolution, the overall shape of these images possesses a rather flat-topped profile on each emission lobe.
Standard hydrostatic models for almost edge-on disks will produce an intensity with an hourglass or chalice shape, as shown here and in many models presented in Fig.13 in \cite{Duchene2024}. 
The morphology of the images observed for other edge-on disks -- such as for HH30 with HST \citep[e.g.,][]{Burrows1996} and, more recently, in NIRSpec continuum images around 4.65~$\mu$m \citep[see Fig.~2 of][]{Pascucci2024} or 2~$\mu$m \citep[see Fig.~2 of][]{Tazaki2025} where the HH30 hourglass shape of the continuum is clear and departs from \thisdisk -- seems adequately described with such an hourglass shape. Here, it is clear that in the JWST observations of \thisdisk, the intensities extend further out and the resulting observed intensity distributions look more like two ellipsoidal lobes in the NIR. This upper atmosphere of the disk can be seen in the vertical cuts through the disk centre shown in the right panels of Fig.~\ref{Figure_standard_images}, where the observed profiles are wider and extend further out than in the model. 
%
%
%
\begin{figure}[htbp]
\begin{center}
\includegraphics[width=\columnwidth,angle=0]{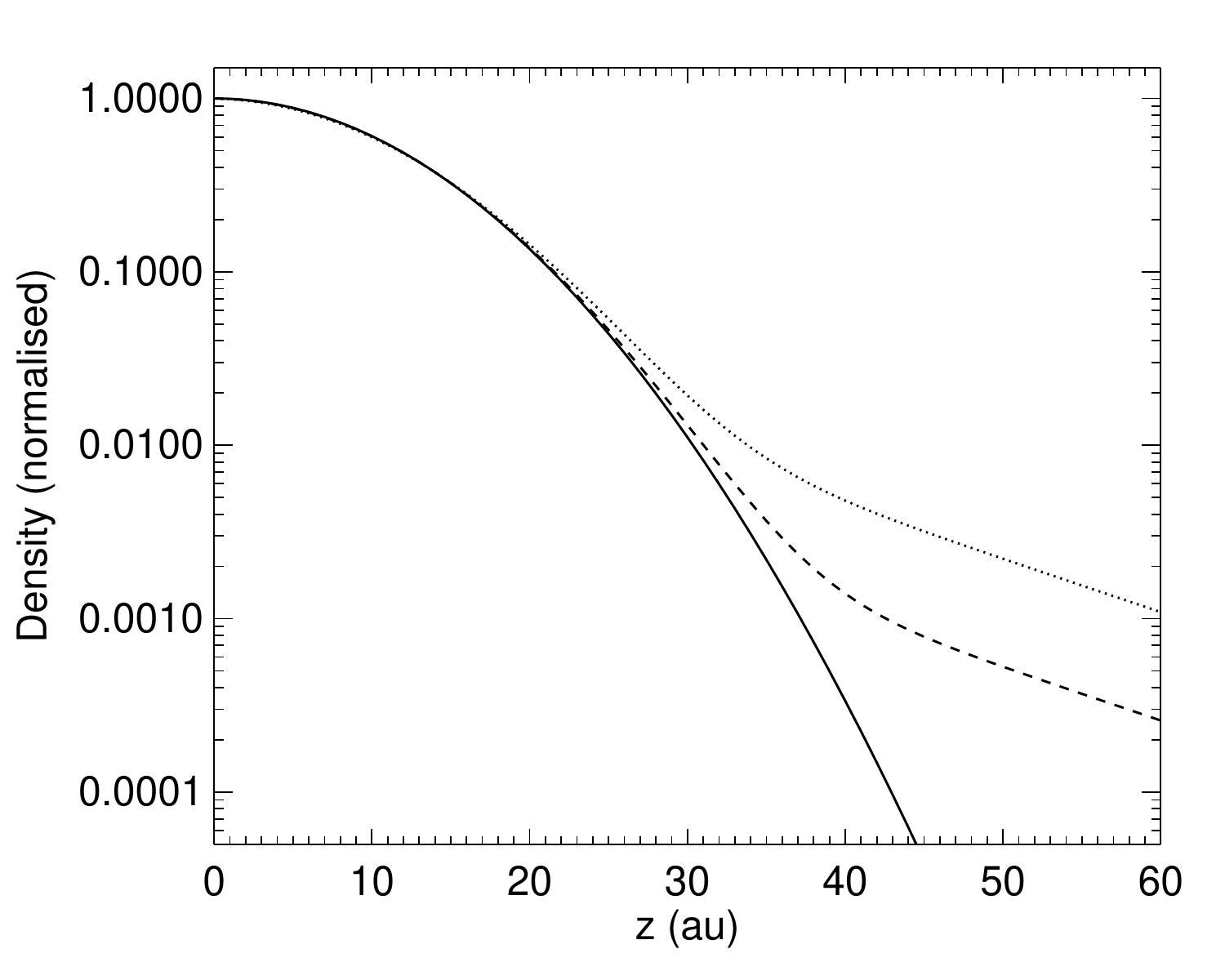}
\caption{Vertical density profile for a scale height $\rm H_0=10$, normalised to the midplane density, considering an isothermal model (solid line) and including an extended atmosphere as shown in equation \ref{eqn_vertical_density_modified} for $\epsilon=2.5$ (dotted line) and $4$ (dashed line), corresponding to an atmosphere contributing above about 3-4$\times$z/h.
The integral of the mass increase above 3 scale heights represents about $1\times10^{-2}$ of the total mass for $\epsilon=2.5$ and about $2\times10^{-3}$ of the total mass for $\epsilon=4$, but strongly affects the radiative transfer in the upper layers.}
\label{Figure_extended_atm}
\end{center}
\end{figure}
%
%
\begin{figure*}[!ht]
\begin{center}
\includegraphics[width=1.3\columnwidth,angle=0]{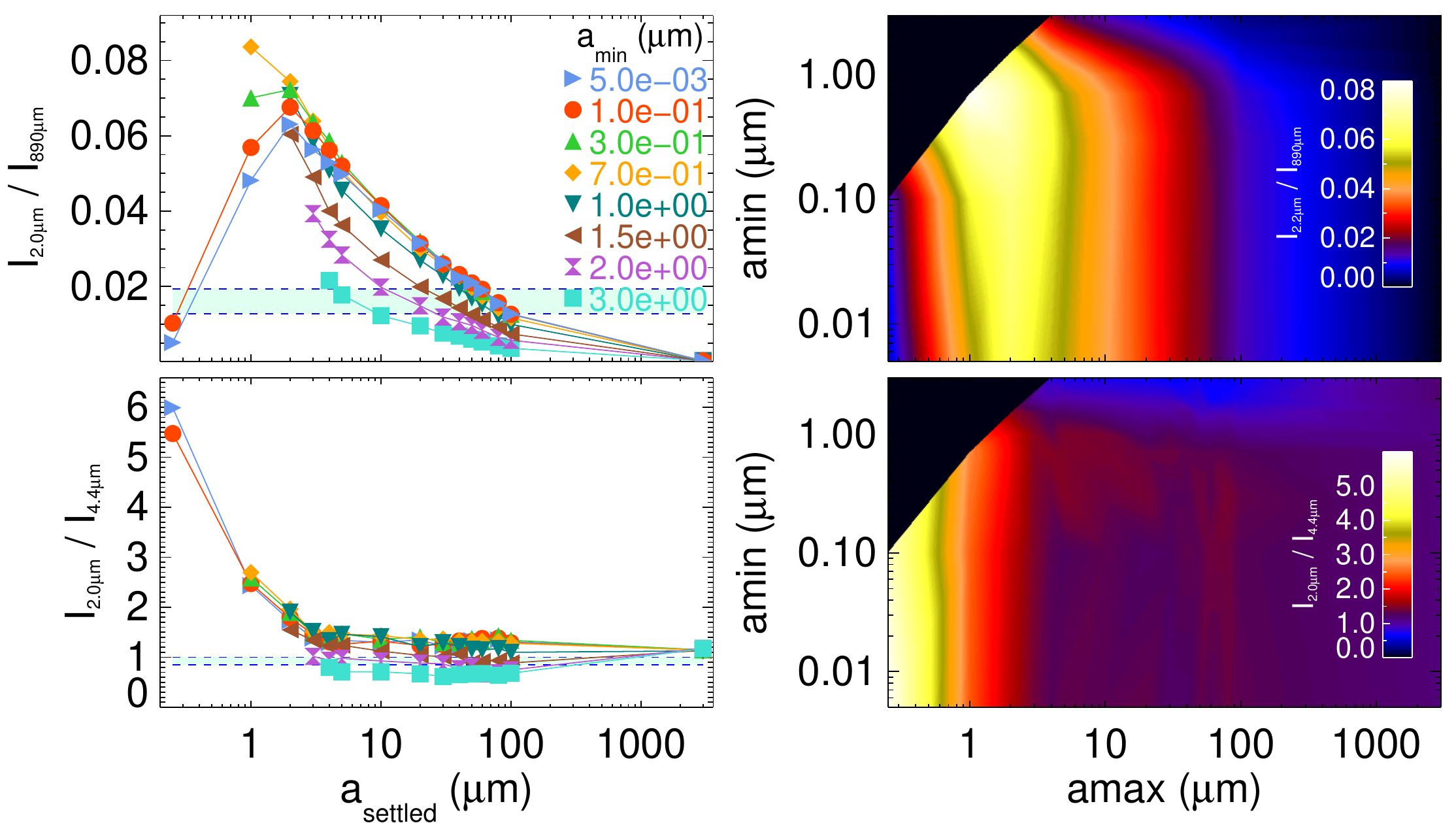}
\raisebox{1cm}{\includegraphics[width=0.7\columnwidth,angle=0]{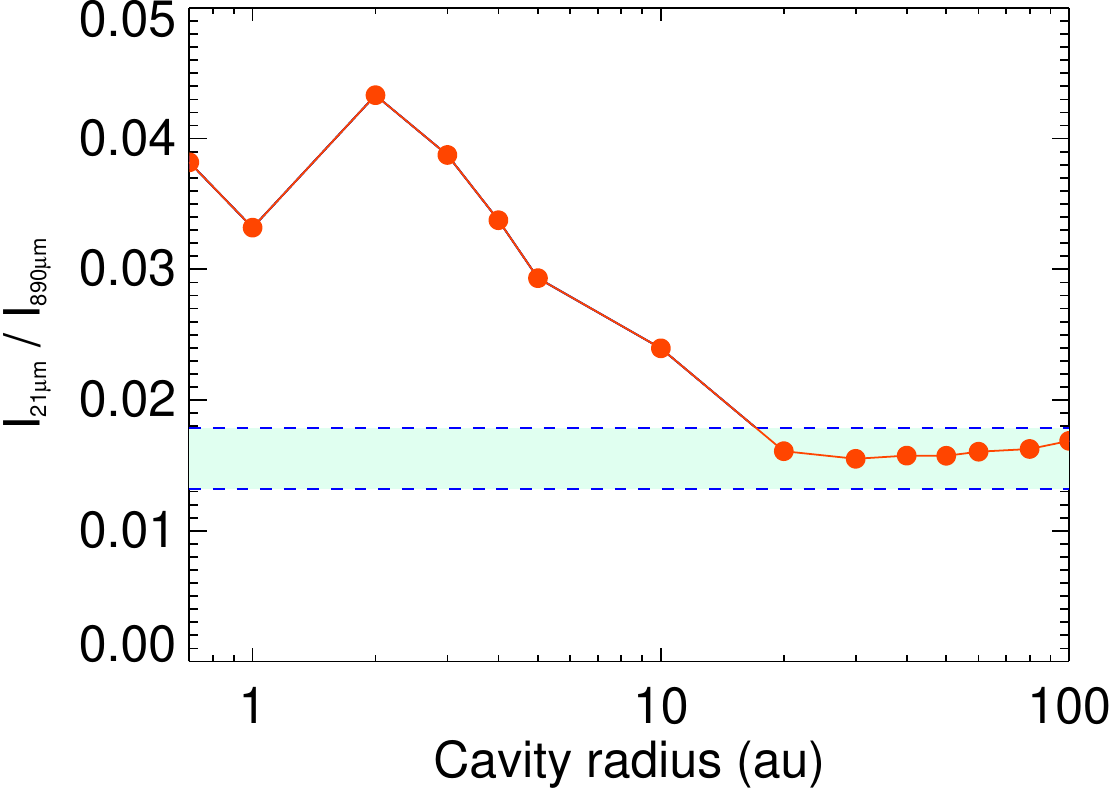}}
\caption{Selected NIR-to-FIR characteristic intensity ratios. Left: Intensity ratios calculated as a function of a$\rm_{settling}$ for the benchmark model. a$\rm_{min}$, defining the evolution of the small and large dust grain distributions, is varied from the MRN value (0.005~$\mu$m) to 3~$\mu$m. 
The shaded regions delimited by dotted lines represent the expected observed intensity ratios, taking into account the variations between JWST observations and previous photometric data points. 
The central panels are 2D representations of the same model results.
Right: Exploration of the effect of cavity radius on the 21 to 890~$\mu$m intensity ratio in the benchmark model, which includes the presence of an inner, partially filled, cavity. A better agreement is obtained for $r_{cav}$ above several tens of au.}
\label{Figure_evolution_ratios_benchmark}
\end{center}
\end{figure*}
%
\subsection{Adding an extended atmosphere to the standard structure}
\label{benchmark_with_extended_atmosphere}
From the results above, and as already concluded by \cite{Duchene2024}, it is evident that some grains are present high in the disk, producing a `veil’ of optical scattered light. This is evidenced when looking at the intensity perpendicular to the disk, as discussed above, with extended wings present well above the disk midplane and especially in the near-infrared. The standard radiative transfer with standard scale heights for small and large grains distributions, well describing the dominant mass fraction of the disk, must thus be modified to include the presence of small amounts of such grains. 
To consider this in the model, a small fraction of gas and grains can be present at a higher scale height than expected for the isothermal vertical density profile. The physical reasons for, and implication of the presence of, a tenuous extended atmosphere will be discussed below.

The vertical density profile of an isothermal disk, as defined using equation \ref{eqn_vertical_density_hydro} implemented in the radiative transfer models described above, results in intensity profiles that are too narrow, i.e., not extended high enough above the disk midplane.  
The observations clearly show a shallow intensity halo that is vertically extended, at several times the local scale heights ($\sim3\times H(r)$)  expected from a purely hydrostatic disk, due to an extended disk atmosphere. At such scale heights above the disk, departures from a pure Gaussian density profile are expected. There exist several physical reasons for that, including among the possibilities: deviation from pure hydrostatic equilibrium in relatively massive disks (and this disk must be massive given its large diameter approaching 1000 au), a non-isothermal temperature profile at large scale height (superheated layers), vertical diffusion, turbulence, the presence of dust ejected as a result of the left-over material from the recently imaged nested jets \citep{Pascucci2024}, or dust sustained at higher heights by the past or present remains of magnetic fields. Such departures have been proposed in previous models and discussed in general terms for many disks \citep{Lee2024, Lebreuilly2023, montesinos2021, lesur2021magnetohydrodynamics, armitage2015, Dutrey2011, Hirose2011, Ciesla2010, chiang1997}. Since, in the case of \thisdisk, we are dealing with an edge-on disk, it is more likely that we can see and probe such an extended upper atmosphere. 
To model the effect of an extended atmosphere, we adopt a simple configuration by slightly modifying the vertical density profile in a way that has been parameterised to empirically fit some of the protoplanetary disk models cited just above. We modify equation~\ref{eqn_vertical_density_hydro} by replacing it with the following extended atmosphere equation:

\begin{equation}
\rm        \rho^{e.a.}(r, z) = 
\frac{\Sigma(r)}{H(r)\sqrt{2\pi}} \;\;
\frac{\exp\left(- \alpha^2 \right)
+\beta \exp\left(-\alpha\right)}{1+\beta} \;;\; \alpha = \frac{|z|}{\sqrt{2}H(r)}
\label{eqn_vertical_density_modified}
\end{equation}

\begin{figure*}[!ht]
\begin{center}
\includegraphics[width=2\columnwidth,angle=0]{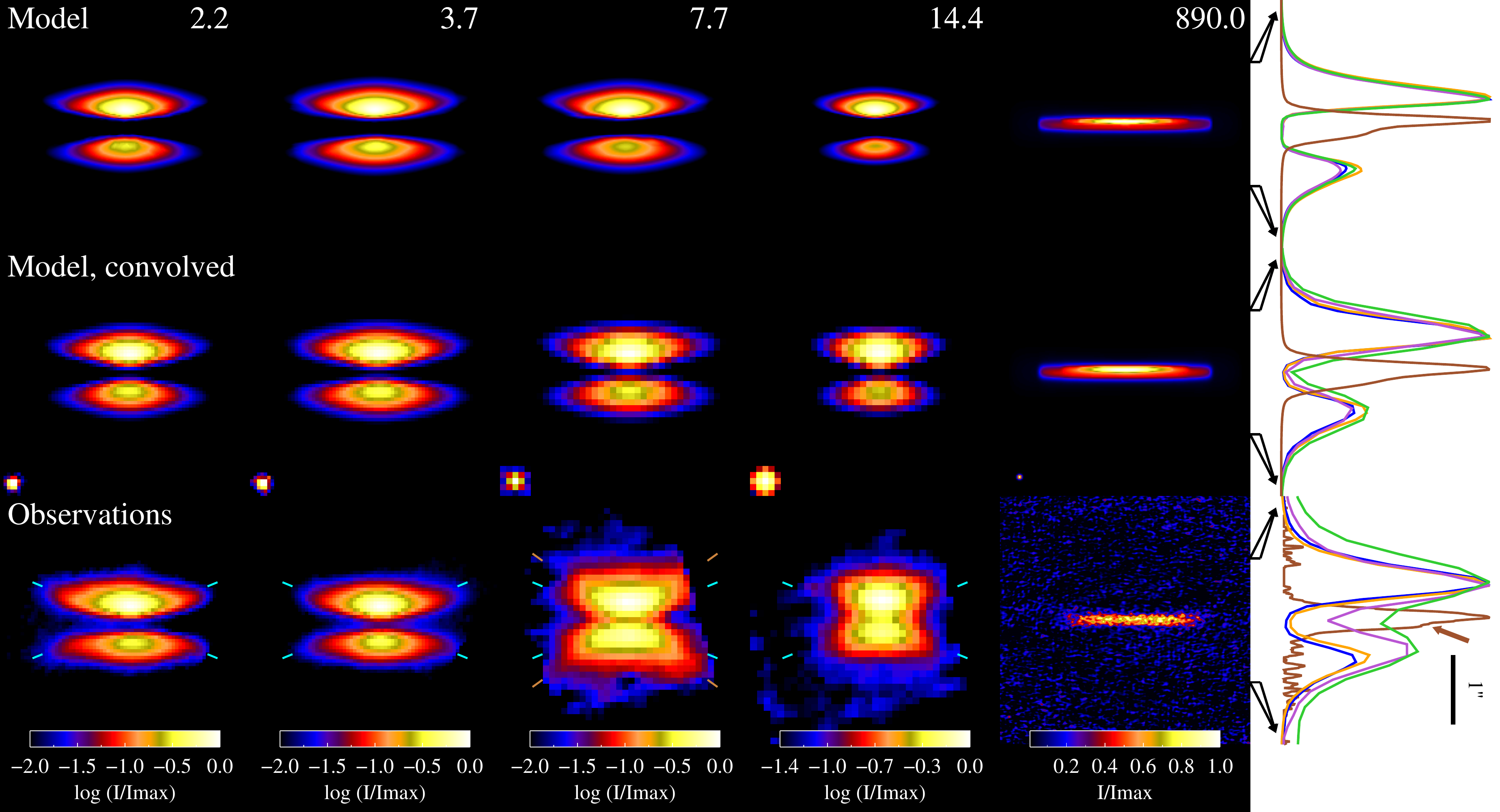}
\caption{Benchmark model, i.e. the `standard' disk model plus an extended atmosphere. 
The nomenclature is the same as for Fig.\ref{Figure_standard_images}.
The presence of an efficient scatterer rather high up in the disk in this extended atmosphere allows for a better reproduction of the vertical distribution of observed intensities. See text for details.
A version of this same figure with a uniform red temperature colour table is presented in the appendix.}
\label{Figure_extended_atmosphere_images}
\end{center}
\end{figure*}

The first term is the classical isothermal hydrostatic density profile from equation~\ref{eqn_vertical_density_hydro}, while the second term produces a tail in the density profile corresponding to the extended atmosphere.
Parametrising $\rm\beta$ as $\rm \beta = exp(-\epsilon)$, for $\epsilon=\{2,5\}$, the extended atmosphere contribution surpasses the classical isothermal vertical density profile above a few $\rm z/h$, as shown in the middle sketch of Fig.\ref{figures_schematiques}.
Although it represents only a tiny fraction of the total dust mass, this extended atmosphere has a profound effect on the NIR intensity profile of the JWST images and spectra. 
The result of an extended atmosphere model calculation, similar to the standard model, but including the implementation of equation~\ref{eqn_vertical_density_modified} with $\epsilon=2.5$, is shown in Fig.\ref{Figure_extended_atmosphere_images}, and is referred to as the `benchmark model' hereafter. This $\epsilon=2.5$ value allows the description of the vertical extent of the observed disk emission, especially in the NIRSpec wavelength range. As shown in Fig.~\ref{Figure_extended_atm} it implies mobilising about or less than 1\% of the disk dust surface density above three scale heights.
Using this benchmark model including an extended atmosphere, we will try to reproduce the SED, the ice features' spectral profiles, and the observed images in the following sections.
%
%
%
\begin{figure}[!ht]
\begin{center}
\includegraphics[width=\columnwidth,angle=0, trim={0cm 0cm 0cm 0}, clip]{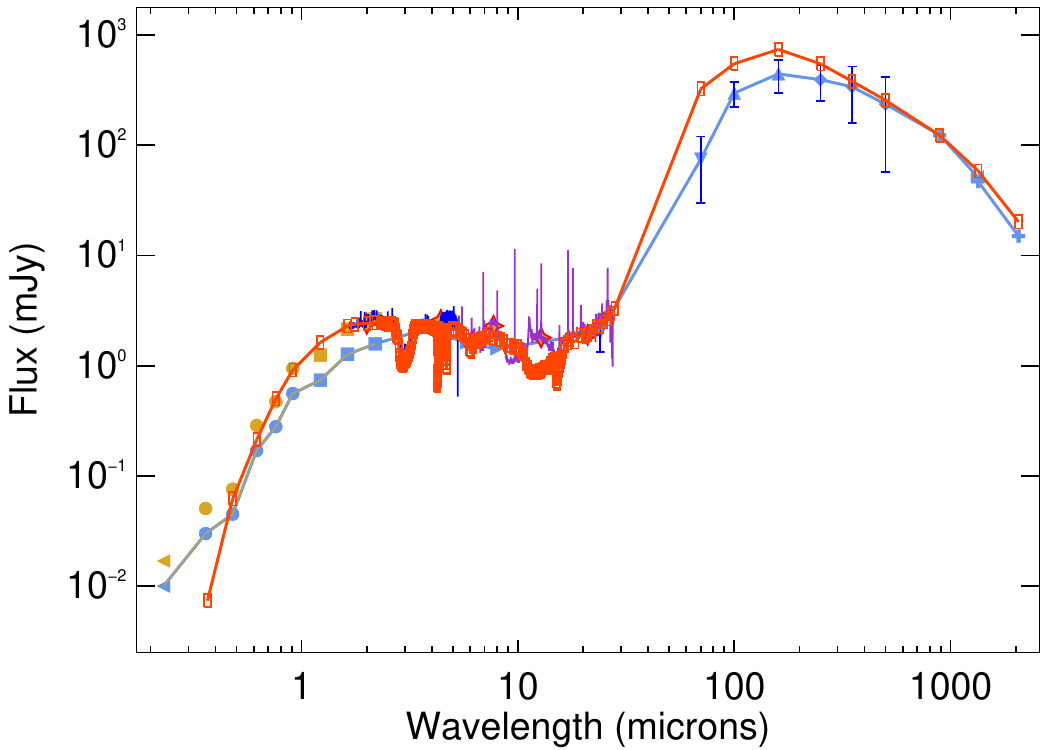}
\caption{Benchmark model spectrum (orange) overplotted on the spectral energy distribution data for \thisdisk, normalised to the observed average flux in the NIRSpec range. Blue photometric points are taken from Table 4 of \cite{Duchene2024}. Orange symbols correspond to the $\leq 2.2~ \mu$m photometric points adjusted by a factor to match the JWST flux, taking into account the known potential variability of the source (see text for details). Error bars (3$\sigma$) are given for the long wavelength photometric points from Spitzer/MIPS, Herschel/PACS and Herschel/SPIRE, retrieved from published point source catalogues.
}
\label{Figure_extended_atmosphere_SED}
\end{center}
\end{figure}
%
%
\begin{figure*}[!ht]
\begin{center}
\includegraphics[width=2\columnwidth,angle=0 , trim={0 0cm 0cm 0}, clip]{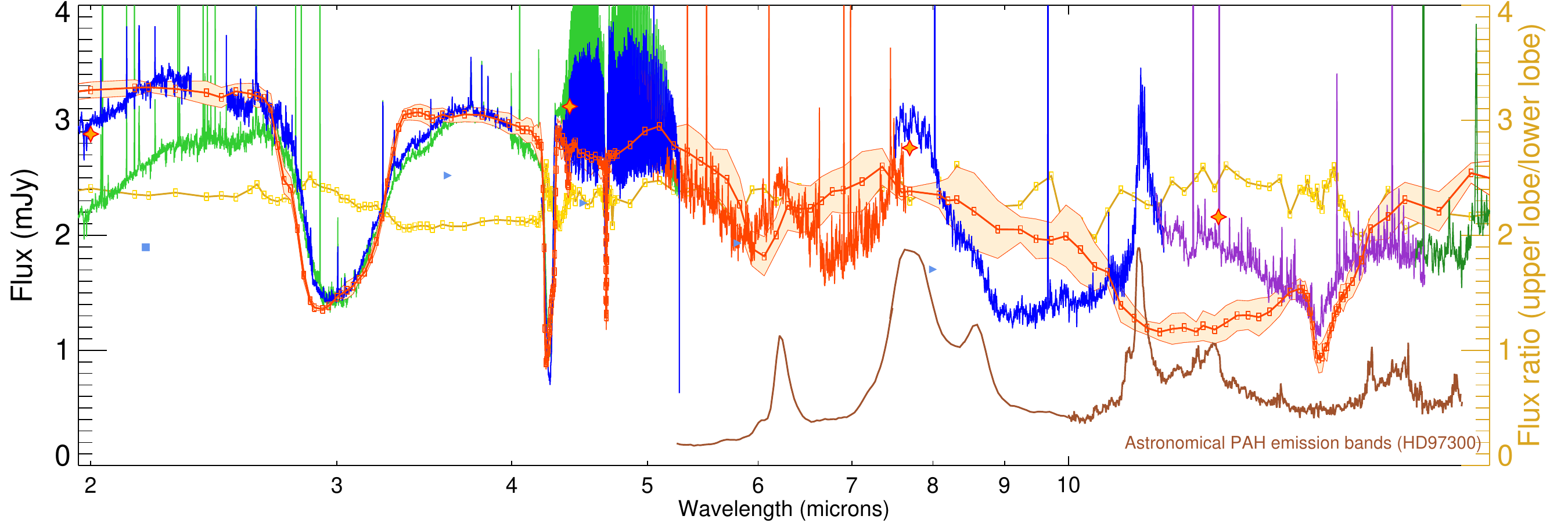}
\caption{Zoom of the benchmark model spectrum (orange) and estimated uncertainties on the calculation (light orange filled region) in the JWST spectral range, overplotted on the NIRSpec-MIRI combined spectrum. The astronomical PAHs emission from the HD97300 disk is shown in order to delineate the regions where such features contribute to the spectrum (these emission features are not explicitly included in the model).}
\label{Figure_extended_atmosphere_spectres_obs}
\end{center}
\end{figure*}
%

\begin{figure*}[!ht]
\begin{center}
\includegraphics[width=0.66\columnwidth,angle=0]{Figures/image_nirspec_od_3.0_sur_3.7.pdf}
\includegraphics[width=0.66\columnwidth,angle=0]{Figures/image_nirspec_od_4_26_sur_4_32.pdf}
\includegraphics[width=0.66\columnwidth,angle=0]{Figures/image_nirspec_od_4_67_sur_4_72.pdf}

%
\includegraphics[width=0.66\columnwidth,angle=0]{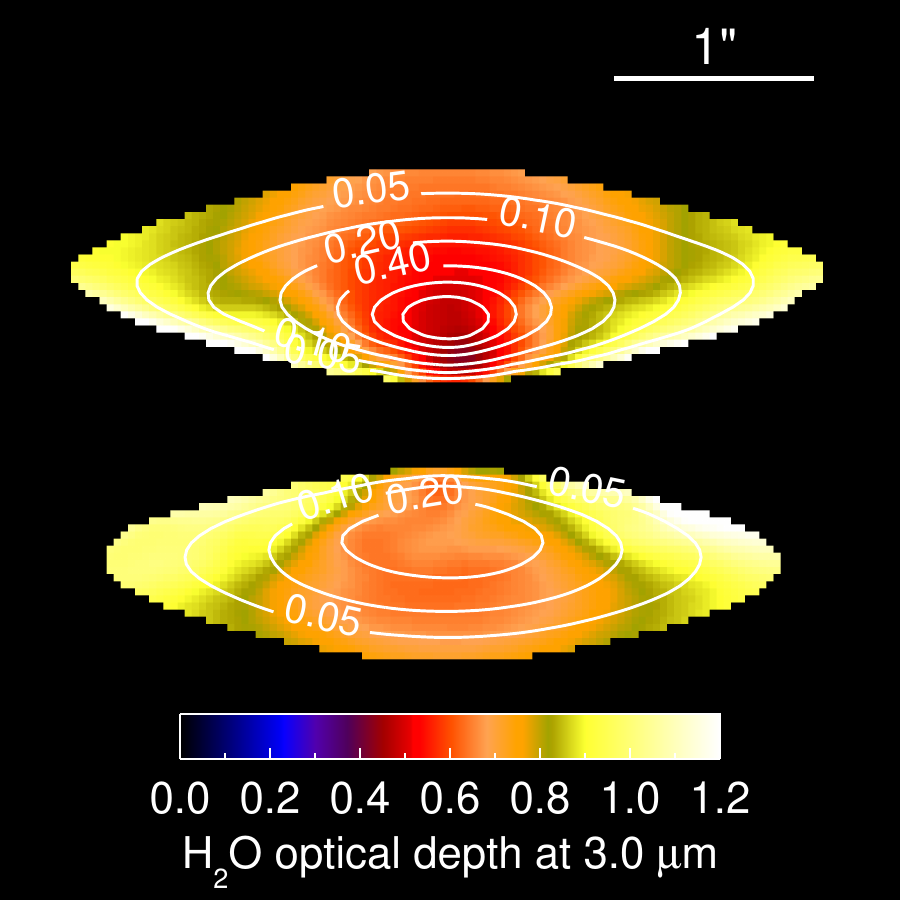}
\includegraphics[width=0.66\columnwidth,angle=0]{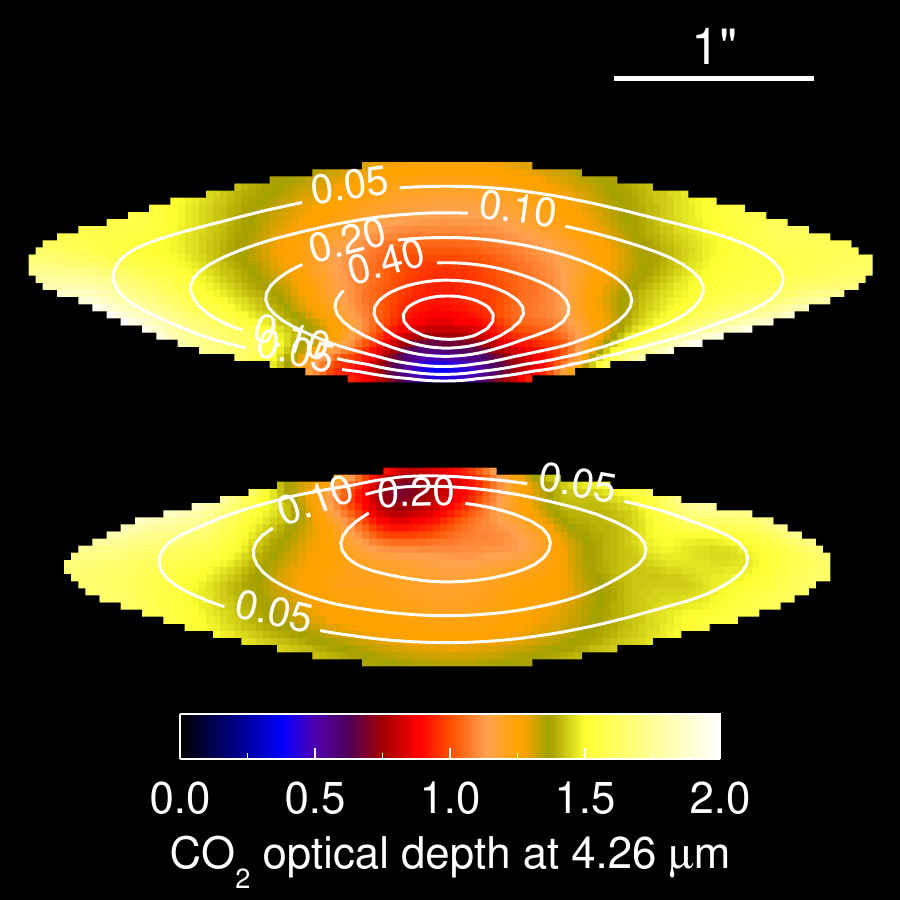}
\includegraphics[width=0.66\columnwidth,angle=0]{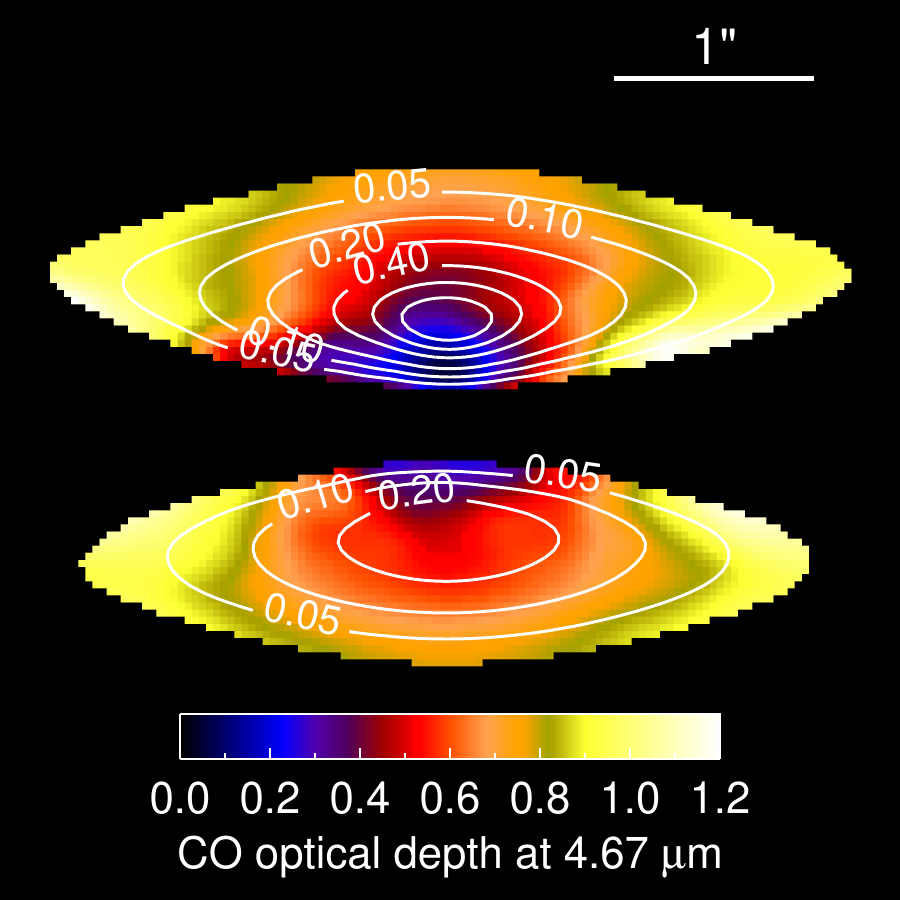}
\caption{Upper row: reproduction of Fig.\ref{Figure_optical_depth_ices_obs}, observed ice optical depth mapped images. Lower row: Benchmark model optical depth images of the disk in H$_2$O, CO$_2$ and CO, as derived at 3.0, 4.26, 4.67~$\mu$m calculated against the reference continuum intensities taken at 3.7, 4.32, 4.72~$\mu$m, respectively. Over-plotted white contours show the flux intensities at the band centres, normalised to their maximum in the map. The last contour is at 5\% from the maximum which is approximately the limit achieved in the observations.}
\label{Figure_optical_depth_ices_obs_sim}
\end{center}
\end{figure*}
%
\subsubsection{FIR/NIR SED provides global constraints on grain sizes}
The extensive probe of all the possible parameters to be explored is a tricky aspect of disk modelling at such a level of detail. The calculation time required for the  production of our RADMC3D models under full scattering radiative transfer being prohibitive, due to the need for high photon statistics considering the high level of scattering involved, we used a coarse grid of parameters for the size distributions with minimum size a$\rm_{min}$ ranging from 0.005~$\mu$m to 10~$\mu$m, with the maximum size for this distribution, a$\rm_{settling}$ from 0.25~$\mu$m to 100~$\mu$m. The upper size range is fixed to 3 mm for the largest grains. The scale height at the reference radius $\rm r_0 = 100$~au was varied between 6 and 10 au. The flaring index $h$ was explored between 1.05 and 1.25.
In order to narrow the parameter space to be explored in models, the intensity ratios in the observed SED of the source, i.e., comparing millimetre emission to the near infrared, as well as the slope in the mid-IR, provide constraints on the underlying and required dust distributions.\\
Based on the available photometric data, \cite{Duchene2024} explored six different ice-free radiative transfer models, concluding that grains up to 10~$\mu$m in size are fully coupled to the gas up to the disk's surface layers.
In agreement with these results, we need a settled component of dust grains, and we explore hereafter the allowed range for this settling size. 
In addition, a better agreement is obtained to the ratio between the $20-50~\mu$m range fluxes and the mm flux if we include a cavity of a few tens of au. A 40 au cavity is close to the minimum cavity size necessary to match the 21~$\mu$m to 890~$\mu$m intensity ratio, with much lower cavity sizes producing an emission ratio in excess of what is observed.
We explore in more detail the evolution of the intensity ratios with different models, including ices, and systematically varying a$\rm_{min}$ and a$\rm_{settling}$ while setting the other parameters to values close to the midpoint of the previously explored parameter grid. 
This is shown in Fig.~\ref{Figure_evolution_ratios_benchmark} for several models we calculated, concentrating on and starting from the benchmark model, then varying the a$\rm_{min}$ and a$\rm_{settling}$ values.
A number of general trends have emerged from this parametric study. 
The essence of the results obtained is that the 2~$\mu$m over 890~$\mu$m intensity ratio first increases when a$\rm_{settling}$ increases and then decreases. For a$\rm_{min}$ larger than 2$\mu$m, the decrease is pronounced. 
All models explored with a$\rm_{settling}$ below a few microns failed to produce enough scattering intensity in the near infrared with respect to the millimetre as observed. The same is true for distributions with a$\rm_{settling}$ above about 50~$\mu$m. 
Models with a$\rm_{settling}$ below a few microns also fail to produce the 2.0 to 4.4~$\mu$m ratio observed, with, for example, MRN-like distribution producing too steep a slope. The combination of just these two intensity ratios already restricts the parameters space.\\ 
Among the numerous parameters, based solely on the spectral energy distribution, the combinations of the NIR, Mid-IR, and FIR ratios suggest that with a model comprised of two size distributions and including settling, the dust settling cutoff size should lie in the few tens of micrometres, whereas the smaller dust distribution minimum size must stay below a few microns, but be above the MRN cutoff as we need an efficient scatterer in the NIR to account for the 2$\mu$m/Flux(ALMA) flux ratio.

A `benchmark' model with these characteristics is discussed hereafter both in terms of distribution of intensities in images and spectroscopic ice band depths and profiles.

\begin{figure*}[!htb]
\begin{center}
\includegraphics[width=0.775\columnwidth,angle=0]{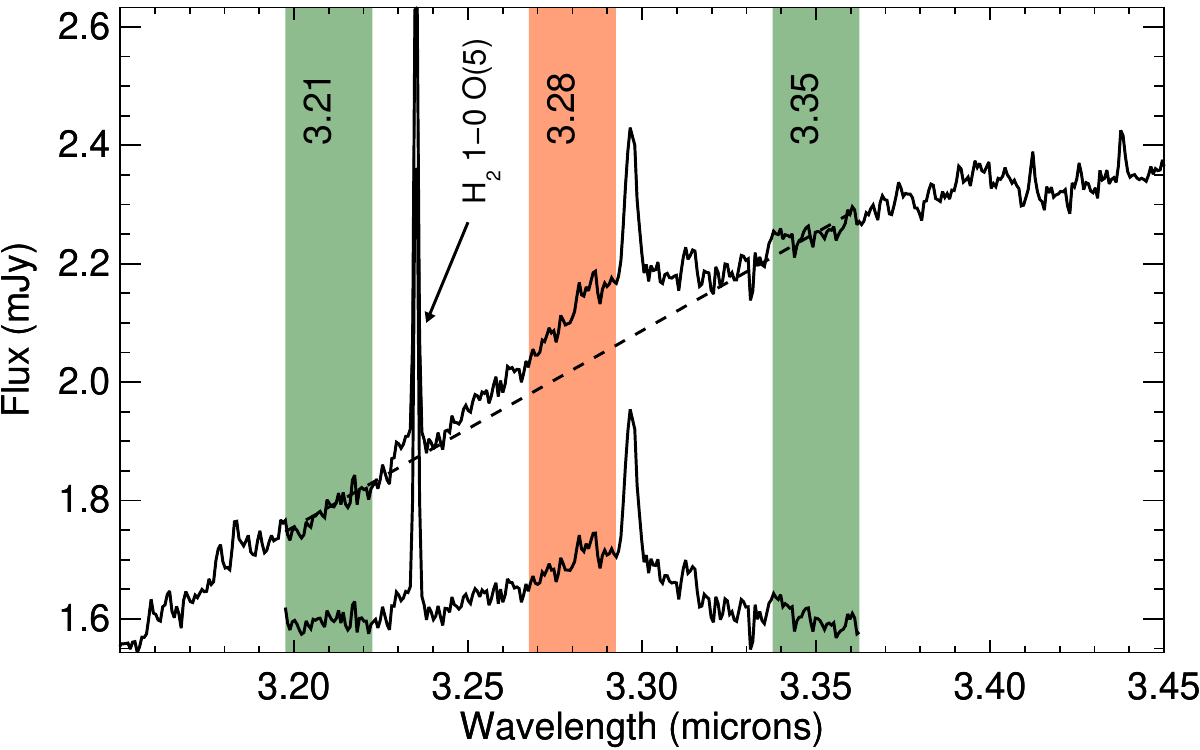}
\includegraphics[width=1.225\columnwidth,angle=0]{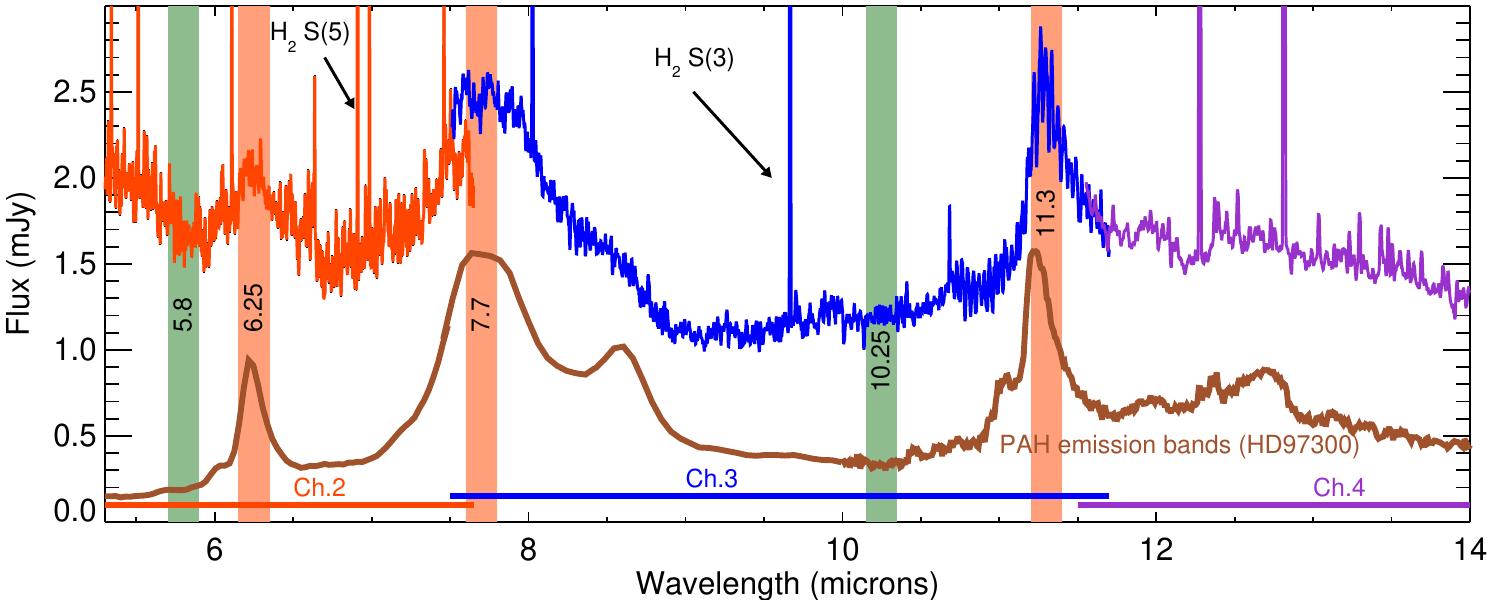}
\caption{Left panel: NIRSpec-IFU spectrum in the region covering the CH stretching mode of astronomical PAH bands. Highlighted as filled regions (orange, green) are the wavelength spans corresponding to the different spectrophotometric filters we applied for our analysis. Green filters serve as continuum references to determine the continuum to be subtracted from the orange filter, which covers the astronomical PAH feature at 3.28$\mu m$, while excluding the adjacent Pfund $\delta$ H emission line.
Right panel: MIRI-MRS spectrum in each observed channel (Ch2 red, Ch3 blue, Ch4 purple). Highlighted as filled regions (orange, green) are the wavelength spans corresponding to the different spectrophotometric filters we applied for our analysis, and corresponding to specific bands of interest in the spectrum. The orange filters are tailored to measure the main emission bands contributing to the astronomical PAH bands. The green filters are reference wavelengths to record potential contributions from the adjacent astronomical PAH-free continuum. Overplotted in brown is the Spitzer spectrum of HD97300 which displays a similar astronomical PAH bands contribution. 
}
\label{Figure_explication_X_wings}
\end{center}
\end{figure*}

\begin{landscape}
\begin{figure}
\vspace*{3cm}
\includegraphics[width=\columnwidth,angle=0 , trim={0 30cm 0cm 0}, clip]{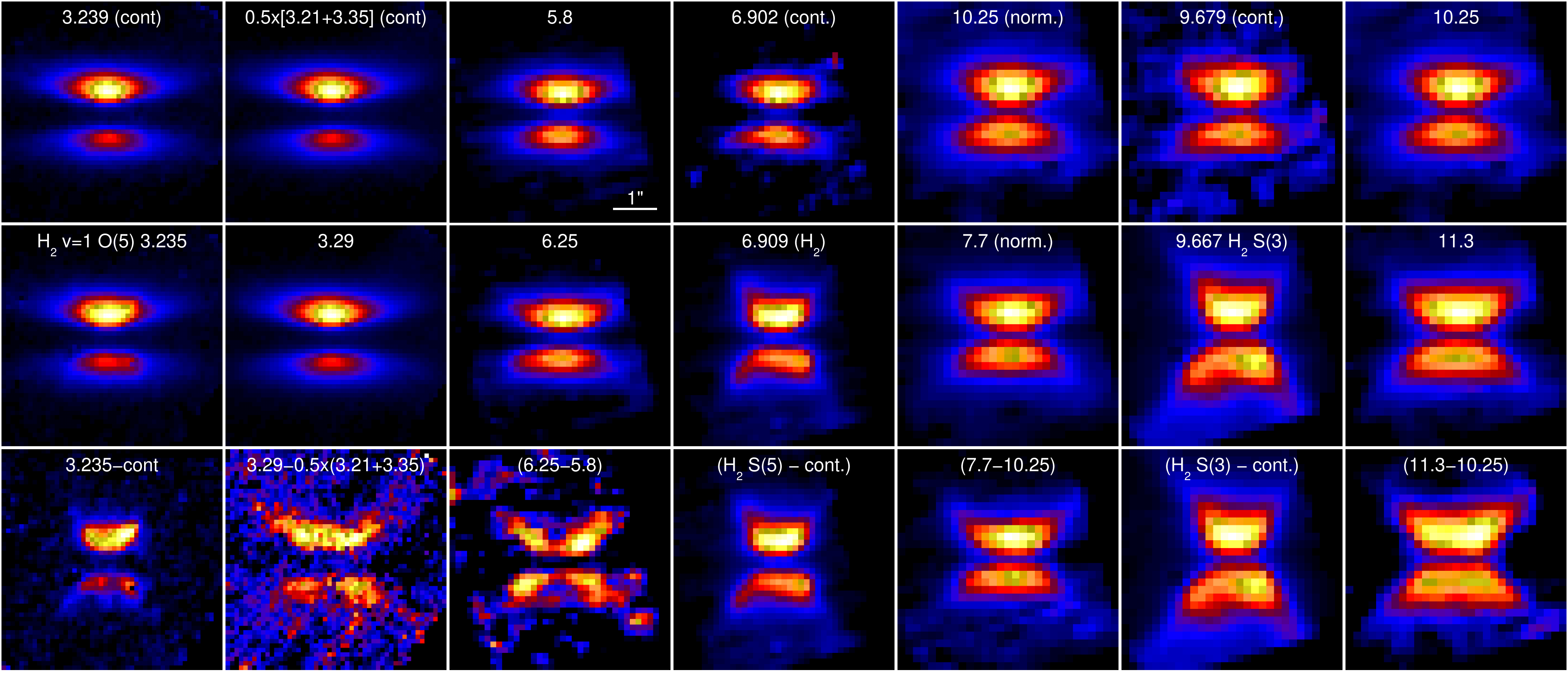}
\caption{
Series of images corresponding to the integration of the observed MIRI-MRS spectra over the newly built-in filters shown in Fig.\ref{Figure_explication_X_wings}. Each image has been normalised to its maximum intensity in order to clearly visualize the distributions of intensities.  The first row corresponds to the `continuum filters' distribution. The second (middle) row contains images of the astronomical PAH bands of interest (6.25,7.7,11.3~$\mu$m) and H$_2$ (6.909, 9.667~$\mu$m) emission lines. The lowest row shows the intensity in bands of interest once subtracted by the adjacent continuum reference intensity, then also normalised to observe the intensity distributions. The upper row shows intensity distributions with globally rounded lobes, whereas each band corresponding to an astronomical PAH band shows a more or less pronounced X shape. The width and opening angle for this X shape are larger than those observed for the H$_2$ emission lines tracing the H$_2$ wind, the latter being more "nested" within this astronomical PAH band X shape.}
\label{Figure_X_wings}
\end{figure}
\end{landscape}

\subsubsection{Images at relevant wavelengths}
The benchmark model fulfilling the characteristics constrained from our simple SED analysis is shown in the first row of Fig.~\ref{Figure_extended_atmosphere_images}, at selected wavelengths spanning the NIR, mid-IR, and the sub-mm from ALMA Band 7. The wavelength selection from the model covers continuum emission wavelengths from the disk, avoiding strong ice bands. 
The second row of images, obtained by convolving the simulated images with the Webb PSF (ALMA PSF for the last image) at each corresponding wavelength can then be compared to the observations presented in the lower row of the figure.
The consequence of the presence of an efficient scatterer rather high up in the disk (above about 3 scale heights), which motivated the addition of an extended atmosphere, is obvious when compared to the standard model images from Fig.~\ref{Figure_standard_images}. It better describes the distribution of intensities as observed with JWST images of the Tau042021 disk.
As shown in the rightmost column -- which compares the intensity profiles of cuts passing through the centre of the disk along the perpendicular axis -- with this extended model, the intensity wings extending up to the upper disk atmosphere are better reproduced than the standard model without the extended atmosphere.
Despite this overall improvement, the central image, at 7.7 $\mu$m, still departs significantly from the model, which suggests it requires an additional component.

The SED calculated from the modelled images is shown in the left panel of Fig.\ref{Figure_extended_atmosphere_spectres_obs}, and a detailed zoom of the NIRSpec-MIRI range is shown in the right panel. The global agreement is good, albeit with a remaining excess still present around 100~$\mu$m and below 0.5~$\mu$m. The excess at long wavelength might be due to deviations of the underlying optical constants in the model from the true optical properties of the grains, whereas the excess in the UV-visible below 0.5$\mu$m most probably arises from an accretion driven mechanism that we do not include and thus cannot reproduce in the modelling.

The observed optical depth images first presented in Fig.~\ref{Figure_optical_depth_ices_obs} are now compared in
Fig.~\ref{Figure_optical_depth_ices_obs_sim} to the optical depth images resulting from the extended atmosphere benchmark model. We qualitatively reproduce well both the absolute values of optical depth and their spatial evolution across the disk except for lines of sight directly aligned perpendicular to the pole close to the stellar line of sight. In particular, there is a pronounced high optical depth ridge at the terminator separating the dark disk from the start of bright emission, and a sort of valley with a progressive decrease moving upwards until the disk emission becomes too low to be measurable. 
In the models, as in the observations, the ice band optical depth images extend far from the disk mid-plane and are comparable, in terms of spatial profile, to the adjacent continuum (see Fig.~\ref{Figure_extended_atmosphere_images}). We do not see and measure a sharp frontier for a transition to a sublimating zone for H$_2$O and CO$_2$ ices in the observations, they are present everywhere that we can measure. Only the CO optical depth seems to asymptotically reduce to zero in the outermost parts of the atmosphere at some point. 
This suggests that  ices observed in the upper atmosphere  of the vertical density profile are sufficiently well protected both from VUV photons external to the cloud, and from direct illumination of the source (shielded by the innermost grains ), otherwise they would have been desorbed, or, dragged/entrained into the upper atmosphere (relatively recently).
\subsubsection{CO$_2$ profile}
The profile of the antisymmetric stretching mode of solid CO$_2$ observed in the Chamaeleon cloud against background sources, i.e., in direct extinction \citep[e.g.,][]{McClure2023,Dartois2024}, tends to form an asymmetric profile with a band exhibiting excess extinction in the red wing and -- because the dust grain size at the upper bound of the size distribution reaches micron sizes -- shows an apparent slight flux increase on the left, i.e., towards shorter wavelengths. In contrast, the observation of solid CO$_2$ in protoplanetary disks, observed at a relatively significant degree of inclination, associated with high optical depths and subject to radiative transfer effects, often reveals a profile that appears inverted, with a pronounced blue wing that can extend relatively far from the core of the band absorption (e.g., IRAS04302+2247, \citealt{Aikawa2012}, Fig. 3; \citealt{Pascucci2024}) and such a profile inversion is observed in the case of Tau 042021 \citep[this work,][]{Pascucci2024} and other disks such as HH 48 NE \citep{Sturm2023b, Bergner2024}, FS Tau B and HH 30 \citep{Pascucci2024}.

The ice band profiles in this work, including the pronounced CO$_2$ feature blue wing extinction, provide constraints on the range of parameters for the models. This is especially true for the CO$_2$ band, because, for a solid phase absorption band, it is intense and narrow. Additionally, it falls in a region of the spectrum fairly free of competing features. 
The inversion of the CO$_2$ profile, which is reproduced in our benchmark model, as shown in the model overplotted over the observations in the right panel of Fig.\ref{Figure_extended_atmosphere_spectres_obs} provides a stringent spectral constraint on the models.
\subsection{Testing the presence of a wind containing astro-PAHs}\label{sec:wind}
\subsubsection{Observational constraints}
As mentioned above, and highlighted by the study of the 7.7$~\mu$m emission, the `benchmark' model, i.e. the model with extended atmosphere, is (in some of the images) insufficient to explain the spatial distribution of the intensity. An X shape is observed to emerge from, and extend beyond, the smooth and progressive evolution of the disk images attributable to the (icy) dust grain bands and continuum alone. 
A peculiar X-shaped feature was already observed in broadband NIRCam and MIRI imaging at 7.7 and 12.8 $\mu$m and discussed by \cite{Duchene2024}. This feature was absent from their other images at 2.0, 4.4 and 21.0~$\mu$m.
An H$_2$ wind has previously been identified from line images constructed for H$_2$ rovibrational transitions from spectral observations with MIRI and NIRSpec, and discussed in \cite{Arulanantham2024} and \cite{Pascucci2024}, respectively. The H$_2$ transitions, because of their high fluxes, also contribute to the previous wide-band sensitive images, as well as other potential components, and this is where constructing narrower band images from the IFU spectral data can help in discriminating the carriers.
Some IFU images in the mid-infrared require the existence of an additional component to explain the observations.
As these structures could follow the wavelengths characteristic of the emission linked to the astronomical PAH bands \citep[e.g.,][]{Chown2024,vanDiedenhoven2004,Peeters2002,Leger1984,Allamandola1985}, whose prominent bands are expected around 3.3, 6.2, 7.7, 11.3~$\mu$m, we produced filtered images in dedicated bands corresponding to these features to demonstrate this. 
Guided by the MIRI integrated disk spectrum of \thisdisk~as shown earlier in Fig.~\ref{Figure_gas_contributions}, we have averaged several spectral images in a wavelength band around positions crucial for determining the contribution of the astronomical PAH bands, and did the same for nearby wavelengths to sample the adjacent continuum disk emission outside these bands. We also selected nearby wavelength ranges with intense H$_2$ transitions to compare them to the expected structure associated with the already identified H$_2$ wind.
The `line filter' wavelength ranges are shown in the upper panels of Fig.~\ref{Figure_X_wings}, for NIRSpec (left) and MIRI (right) while the corresponding images constructed from these filters are shown in the lower part of Fig.\ref{Figure_X_wings}. 
The bands of interest to be cross-compared, on which these filters have been positioned, are at 3.28, 6.25, 7.7, and 11.3~$\mu$m to monitor the astronomical PAHs emission bands, and 3.325, 6.909, 9.667~$\mu$m for H$_2$, corresponding in the case of the latter to the H$_2$ 1-0 O(5), H$_2$ 0-0 S(5) and 0-0 S(3) transitions, respectively. Continuum images have been prepared for NIRSpec from filters at 3.21 and 3.35~$\mu$m (to estimate the continuum baseline to be subtracted from the 3.28~$\mu$m astronomical PAH band corresponding to the CH stretching mode). For MIRI, filters were placed at 5.8 and 10.25~$\mu$m to monitor the continuum outside, but sufficiently close to, the astronomical PAH bands centred at 6.25, 7.7, and 11.3~$\mu$m. In the MIRI range, the spectrum of astronomical PAH emission bands in another disk (HD97300) observed with Spitzer (retrieved from the Spitzer archive as explained previously) is plotted below \thisdisk~to show that the chosen continuum wavelength positions fall in minima with respect to the astronomical PAH bands. These filters are used to estimate and properly subtract the other components in the disk (i.e. the ice features and dust continuum emission) contributing to the astronomical PAHs images.
The bottom row of Fig.~\ref{Figure_X_wings} shows the distribution of intensities for continuum-subtracted astronomical PAH bands 
 observed at 3.28, 6.25, 7.7, and 11.3~$\mu$m. There is evidence of an X shape in each of the three bands. This shape is already seen is the raw images (i.e. not yet continuum subtracted) at 6.25, 7.7, and 11.3~$\mu$m, although the contrast is much higher when the continuum filter contributions are removed.
 At 3.28~$\mu m$, the contrast is far less intense, and the PAH astronomical band does not appear to be present in the raw image, but becomes evident upon continuum subtraction.
When the images fall at wavelengths corresponding to astronomical PAH bands in the MIRI range, the shape of the emission contrasts with the rounded, less extended, shape of the ice and/or dust continuum, confirming that the astronomical PAHs are the most probable carriers of this X-shaped emission.

Another result of this analysis, obtained by comparison with the H$_2$ emission bands, is that this astronomical PAHs-related X-shaped distribution of intensity is much more angularly extended than
the H$_2$ wind.
The width and opening angle for the X shape seems more open than that observed for the H$_2$ emission lines tracing the H$_2$ wind, the latter appearing to be nested in this astronomical PAH bands' X shape.

\begin{figure}[!ht]
\begin{center}
\includegraphics[width=0.325\columnwidth,angle=0]{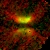}
\includegraphics[width=0.325\columnwidth,angle=0]{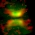}
\includegraphics[width=0.325\columnwidth,angle=0]{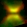}
\caption{Astronomical PAH band-H$_2$ wind composite images. Left: composite of the 3.28~$\mu$m astronomical PAH band (in red) and the 3.235~$\mu$m H$_2$~$1-0$~O(5) line (in green). Middle: composite of the 6.25~$\mu$m astronomical PAH band (in red) and the 6.909~$\mu$m H$_2$~S(5) line (in green). Right: composite of the 11.3~$\mu$m astronomical PAH band (in red) and the 9.667~$\mu$m H$_2$~S(3) line (in green).}
\label{Figure_X_wings_composites}
\end{center}
\end{figure}

To demonstrate this more clearly, we constructed three colour composite images merging astronomical PAH bands (in red) with nearby H$_2$ emission lines (in green), as presented in Fig.~\ref{Figure_X_wings_composites}. The pairs are: the 3.28~$\mu$m astronomical PAH band and the 3.235~$\mu$m H$_2$~$1-0$~O(5) line (left),
the 6.25~$\mu$m astronomical PAH band and the 6.909 H$_2$~S(5) line (centre), and the 11.3~$\mu$m feature and the 9.667 H$_2$~S(3) line (right).
These images show that the bands associated with astronomical PAH carriers seem to form a conical envelope that encloses the H$_2$ emission, such that the transitions of these two carriers form an imbricated stepped system. If we add this to the conclusion drawn by \cite{Duchene2024}, that the warm molecular layer as traced with $\rm^{12}CO$ would be located between the 5--8 $\mu$m scattered light tracing the disk surface and the X-shaped feature, then going perpendicular to the conical structure walls we would cross frontiers delimited by a stratification of a nested wind of H$_2$/astronomical PAHs/warm $\rm^{12}CO$, before reaching the extended atmosphere comprised of gas and dust grains.

The high sensitivity of the broadband MIRI 770W and 1280W images presented in \citet{Duchene2024} reveals a more spatially extended, weak emission. This could also be related to the PAHs identified here via their main vibrational emission bands (3.3, 6.2, 7.7, and 11.3~$\mu$m) thanks to the combined MIRI and NIRSpecIFU spatially resolved spectroscopy. Such an attribution would require spectroscopic information to rule out the contribution of other emission features falling within the 770W and 1280W imaging filters, most notably the broad continuum emission and the sharp molecular H$_2$ and atomic HI and [NeII] lines (see Fig. 6 in \citet{Duchene2024}; \citep{Arulanantham2024} and our Fig 12). The next generation of space-based infrared telescopes would benefit from having some narrow imaging filters centered on the most intense mid-IR PAH features as well as on their adjacent continua.
%
%
%
\begin{figure*}[!ht]
\begin{center}
\includegraphics[width=2\columnwidth,angle=0]{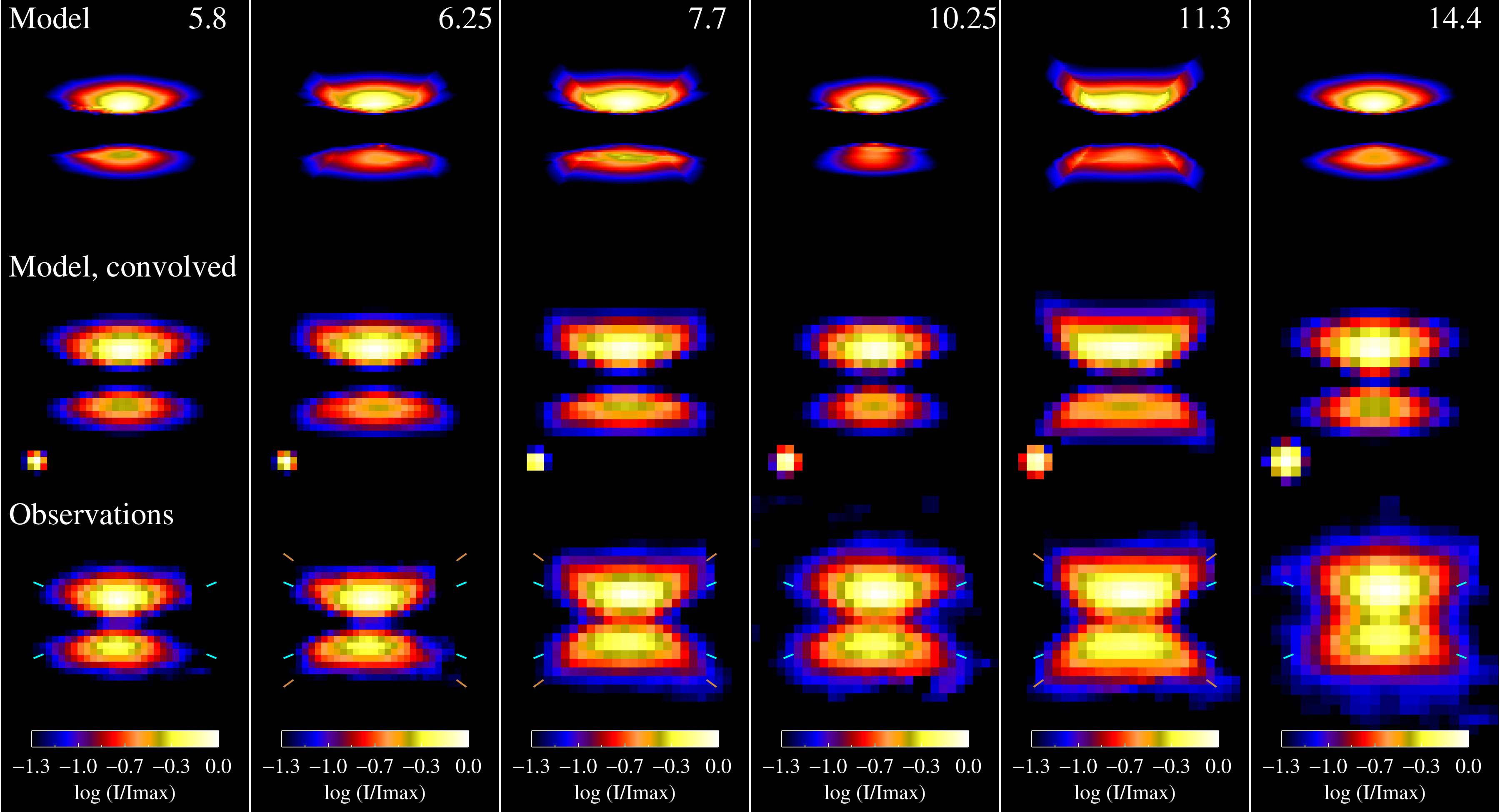}
\caption{An extension of the benchmark model to test the presence of an astro-PAH wind component. (upper) Model images at full resolution for selected wavelengths spanning the mid-infrared for bands including astronomical PAH band contributions (6.25, 7.7, 11.3~$\mu$m) or bands representing the adjacent continuum (5.8, 10.25, 14.4~$\mu$m), for comparison. (middle) Model
images once convolved with the JWST PSF. Note the alternance of a more rounded shape for continuum-dominated bands and a chalice/hourglass shape for those with astronomical PAHs contributions. (lower row) Observed MIRI images. The colorbars indicate the intensity levels, normalised
to the maximum intensity for each image, on a log scale. A version of this same figure with a uniform red temperature color table is presented in the appendix.}
\label{Figure_X_wings_transfer}
\end{center}
\end{figure*}
%
%
\subsubsection{Simple model of a wind with astro-PAHs}
\label{wind_model_containing_astro_PAH}

We will show that the observations can be explained to first order by adding a simplified astro-PAH model to the benchmark disk model with an extended atmosphere. The astro-PAHs wind contribution is described here and illustrated in the lower sketch of Fig.~\ref{figures_schematiques}. 
A simple way to include such a wind consists of matter ejected from a conical structure with opening angles from the disk and whose base start from a radius $\rm r_{wind}$ to $\rm r_{wind} + \delta r_{wind}$. The dependence of density on the height of the wind is not easy to set, but models show it should not decrease strongly past the disk surface \citep{Lesur2021, Ray2021}.  
We define 

\begin{equation}
\rm        \rho^{wind}(r, z) = \delta_{wind} \; \frac{\rho[r, z=\zeta H(r)]}{\left[|z|/[\zeta H(r)]\right]^{-\gamma}} \times \left(\frac{r_{wind}}{r}\right)^2\;;\; |z|>\zeta H(r)
\label{eqn_vertical_density_wind}
\end{equation}

where $\zeta$ is a multiplicative factor to the local scale height at which the wind begins, considered as the disk surface for the wind, and the density boundaries ($\rm \rho^{wind}=0$) are defined by $\rm r_{wind}$ to $\rm r_{wind} + \delta r_{wind}$, and the $\theta_{in}$ and $\theta_{out}$ angles. These different components (settling, extended atmosphere, wind) are depicted in Fig.\ref{figures_schematiques}.
$\gamma$ defines the order of the decrease of the density with respect to the scale height. We adopt $\zeta=2$, i.e., the wind density takes its origin at about two scale heights and evolves with $\gamma=1$. 
The $(r_{wind}/r)^2$ factor assumes a dilution factor if the primary source of excitation comes from the stellar photons.

The astro-PAHs spectral grain model was adapted from the small PAH-Carbonaceous Grains description of \cite{Li2001}, in particular the cross-sections and profiles describing the different bands. From the extinction cross-section, we built small grain properties for the carriers constituting this wind. We do not fully treat the stochastic emission properties of such astro-PAHs, since such a treatment is out of the scope of this article. Instead, we make a simplified hypothesis to approach this. We assign the carriers a fixed emission temperature. By doing this we intrinsically assume that we deal both with the same size distribution of small astro-PAHs in the wind and that photoexcitation leads to the same internal equilibrium temperature -- and thus infrared fluorescence emission cascade -- irrespective of where these astronomical PAH carriers are in the wind. The emission intensity is thus modulated by the (radially and vertically decreasing) density of the carriers and the stellar photons dilution factor. The parametrisation from equation~(\ref{eqn_vertical_density_wind}) discussed above, allows us to describe the observations, but additional studies would be needed to further constrain these astronomical PAHs. At this stage we cannot exclude thermal excitation of the carriers.

The result of the extended atmosphere model added by this astro-PAH wind, assuming $\rm \theta_{in}=49^{o}$ and $\rm \theta_{out}=36^{o}$, $\rm r_{wind}=40$~au, $\delta r_{wind}=25$~au, $\zeta=2$ and an equilibrium temperature for the astro-PAHs of 600~K is shown in Fig.~\ref{Figure_X_wings_transfer} and the resulting model spectrum in Fig.~\ref{Figure_X_wings_transfer_model_spectrum}.
The addition of this wind allows us to explain, to first order, the shape of the observed images in the mid-infrared with MIRI, in and out of the astronomical PAHs wavelengths, and with an X shape, which takes on particular prominence when the aromatic bands are crossed.
\section{Conclusions}

We have modelled the full scattering radiative transfer in the very large, almost edge-on, \thisdisk disk, including the evolution of the size and shape of icy dust grains. We added astro-PAHs as a necessary component to explain the observations covering the NIR to mid-IR with NIRSpec and MIRI IFUs. The essential features of the icy dust grains in the \thisdisk~disk are as follows.
\begin{itemize}
    \item The high NIR to FIR intensity ratio in the observations implies efficient scattering by grains 1--3 microns in size, thus grains grow to at least the micron size in this disk. In agreement with previous studies, grains above tens of microns in size settle towards the disk midplane.
    \item With such a close to edge-on disk, the observations are highly sensitive to the vertical density distribution. We have shown that reproduction of the multi-wavelength distribution of intensities in the images of \thisdisk~requires a radiative transfer model with an atmosphere extending above the classical pure isothermal vertical scale height, for scale heights $>>$ 1, unlike in other disks such as HH30.
    \item Deep ice absorptions are observed, indicating the presence of the major ices H$_2$O, CO, and CO$_2$. The spatial distribution of these ices is revealed by constructing optical depth images comparing the absorption features to the adjacent continuum. The ices are determined to be present up to high altitudes above the disk midplane, i.e. above three scale heights. This suggests that ices observed in the upper atmosphere of the vertical density profile are either sufficiently well protected from VUV photons -- both those external to the cloud, and the direct illumination by the source (shielded by the innermost grains) -- to prevent significant desorption, or are replenished on short timescales i.e. have been relatively recently dragged/entrained into the upper atmosphere.
    For H$_2$O and CO$_2$, this vertical frontier is observationally limited only by the sensitivity of JWST, and ices may be present even higher in the disk than probed here. Apart from the very inner part of the disk, the radiative transfer is thus dominated by ice covered dust grains.
    \item These observations underline the presence of a wind containing the carriers of astronomical PAHs. It appears as an X-shaped emission at the wavelengths of the characteristic mid-infrared astronomical PAH bands, i.e. around 3.3, 6.2, 7.7 and 11.3 microns. We provide a framework for interpreting the observations and to show that the astronomical PAH emission is more extended than the ice and dust continuum.
    \item     Images of the astronomical PAH emission and the emission of H$_2$ in Tau 042021 seem to indicate that the spectral bands linked to astronomical PAH carriers outline a conical envelope encasing the H$_2$ emission. This arrangement suggests that the transitions associated with these two carriers form a layered, step-like structure. This is particularly interesting in the context of the conclusion from previous studies proposing that the warm molecular layer traced by $\rm^{12}CO$ is situated between the 5--8~$\mu$m scattered light marking the disk surface and the X-shaped feature i.e. that moving outwards perpendicular to the conical structure’s walls would reveal successive boundary layers.
    These layers would therefore follow a stratified pattern -- consisting of an embedded wind composed of H$_2$, astronomical PAHs, and warm $\rm^{12}CO$ -- before transitioning into the extended atmosphere filled with gas and dust particles.
    Further studies are needed to refine the constraints on the emission characteristics and underlying physics of the astronomical PAHs in \thisdisk.
\end{itemize}

\begin{acknowledgements}
Support for program No. 1751 was provided by NASA through a grant from the Space Telescope Science Institute, which is operated by the Association of Universities for Research in Astronomy, Inc., under NASA contract NAS 5-03127. 
E.D. and J.A.N. acknowledge support from French Programme National `Physique et Chimie du Milieu Interstellaire' (PCMI) of the CNRS/INSU with the INC/INP, co-funded by the CEA and the CNES. 
A portion of this research was carried out at the Jet Propulsion Laboratory, California Institute of Technology, under a contract with the National Aeronautics and Space Administration (80NM0018D0004). 
D.H. is supported by a Center for Informatics and Computation in Astronomy (CICA) grant and grant number 110J0353I9 from the Ministry of Education of Taiwan. D.H. also acknowledges support from the National Science and Technology Council, Taiwan (Grant NSTC111-2112-M-007-014MY3, NSTC113-2639-M-A49-002-ASP, and NSTC113-981 2112-M-007-027). 
M.E.P. acknowledges support by INAF grant within the program Ricerca Fondamentale 2022. 

\end{acknowledgements}

\bibliographystyle{aa}
\bibliography{midas.bib}

\begin{appendix}
\section{Images with a uniform color table}
\label{appendix_A}
%
\begin{figure*}[!ht]
\begin{center}
\includegraphics[width=2\columnwidth,angle=0]{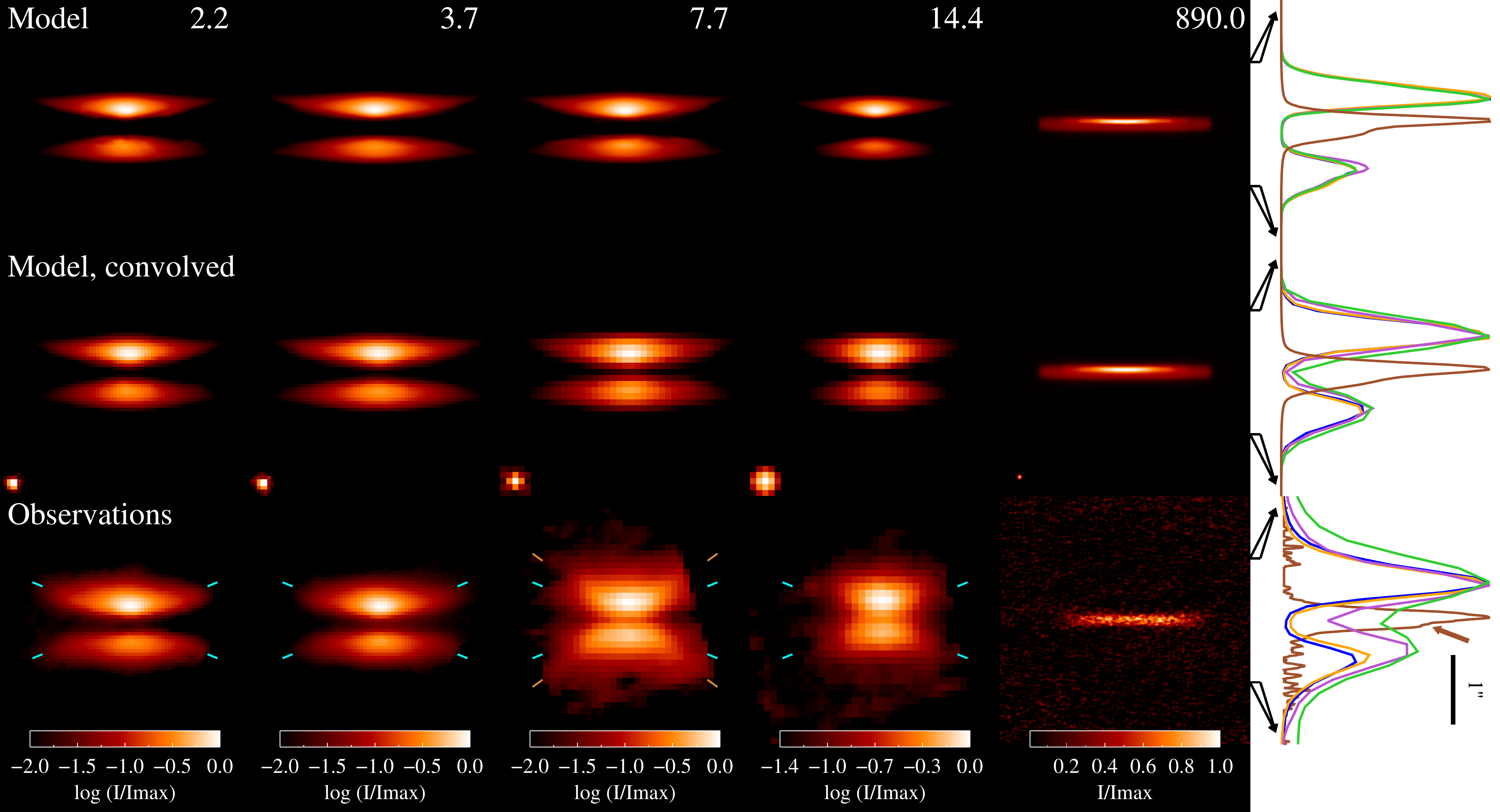}
\caption{Standard model. (upper) Model images at full resolution for selected wavelengths spanning the NIR to mm range. (middle) Model images once convolved with the JWST PSF.
Blue ticks are guides to the small wings observed at about $22.5^{\circ}$, especially in the near infrared, whereas brown tickmarks indicate an angle of $36^{\circ}$ overplotted on the 7.7 $\mu$m image, corresponding to the angle for the X-shape discussed by \cite{Duchene2024}, and close to the H$_2$ wind (semi-opening) angle previously observed in the $35-38.5^{\circ}$ range by \cite{Arulanantham2024, Pascucci2024}. 
The colorbars indicate the intensity levels, normalised to the maximum intensity for each image, in log scale except for ALMA data presented in linear scale. The right panel are cuts along the vertical line through the centre of the disk observations and models at the corresponding wavelengths. The spatial scale is expanded by a factor of two as compared to the images for a clearer view.}
\label{Figure_standard_images_appendix}
\end{center}
\end{figure*}
%
%
\begin{figure*}[!ht]
\begin{center}
\includegraphics[width=2\columnwidth,angle=0]{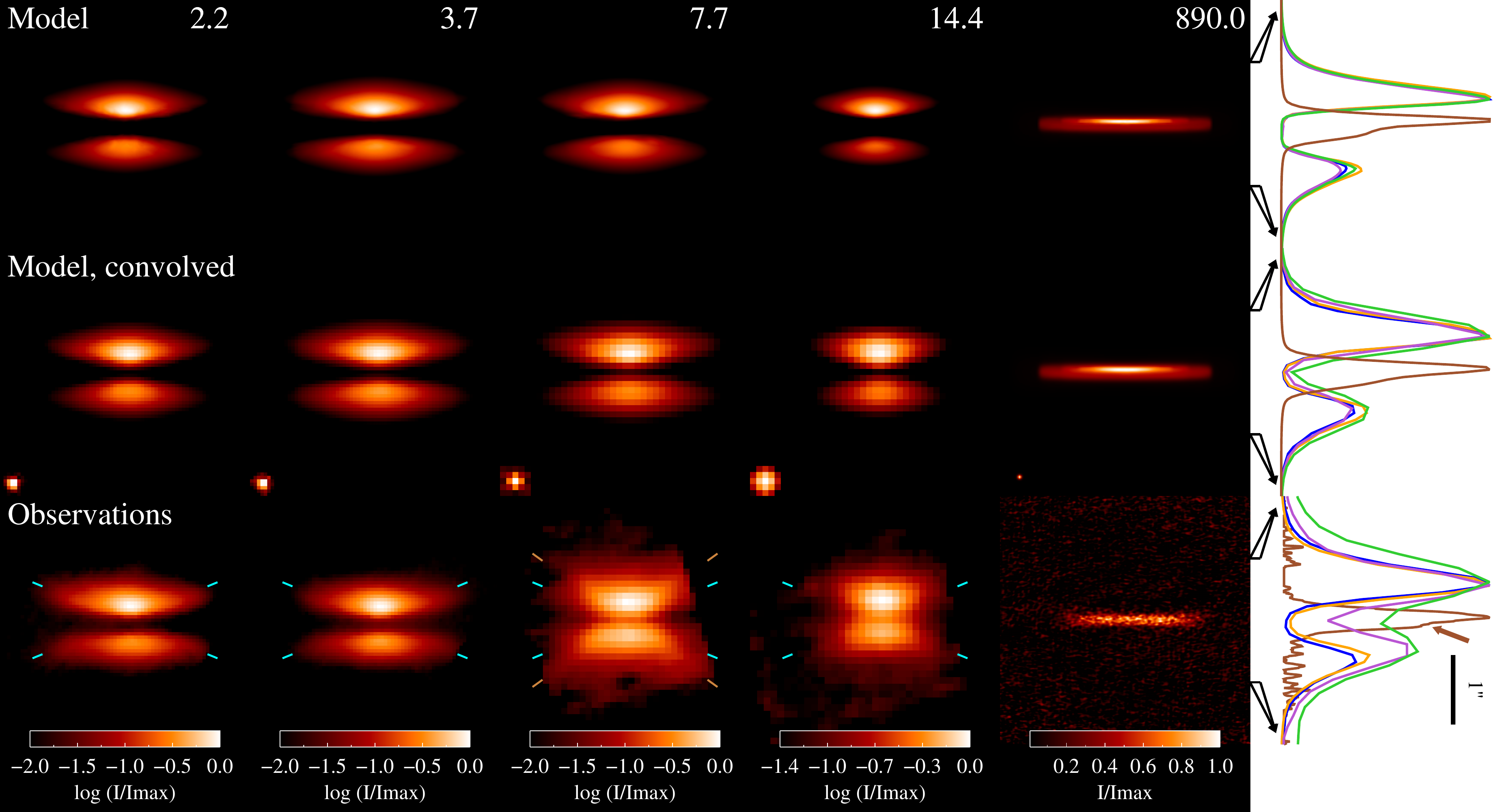}
\caption{Benchmark model. (upper) Model images at full resolution for selected wavelengths spanning the NIR to mm range. (middle) Model images once convolved with JWST PSF.
Blue ticks are guides to the small wings observed at about $22.5^{\circ}$, especially in the near infrared, whereas brown tickmarks indicate an angle of $36^{\circ}$ overplotted on the 7.7 $\mu$m image, corresponding to the angle for the X-shape discussed by \cite{Duchene2024}, and close to the H$_2$ wind (semi-opening) angle previously observed in the $35-38.5^{\circ}$ range by \cite{Arulanantham2024, Pascucci2024}. 
The colorbars indicate the intensity levels, normalised to the maximum intensity for each image, in log scale except for ALMA data presented in linear scale. The right panel are cuts along the vertical line throught the center of the disk observations and models at the corresponding wavelengths. The spatial scale is expanded by a factor of two as compared to the images for a clearer view. A version of this same figure with a single red temperature color table is presented in the appendix.}
\label{Figure_extended_atmosphere_images_appendix}
\end{center}
\end{figure*}
%
%
\begin{figure*}[!ht]
\begin{center}
\includegraphics[width=2\columnwidth,angle=0]{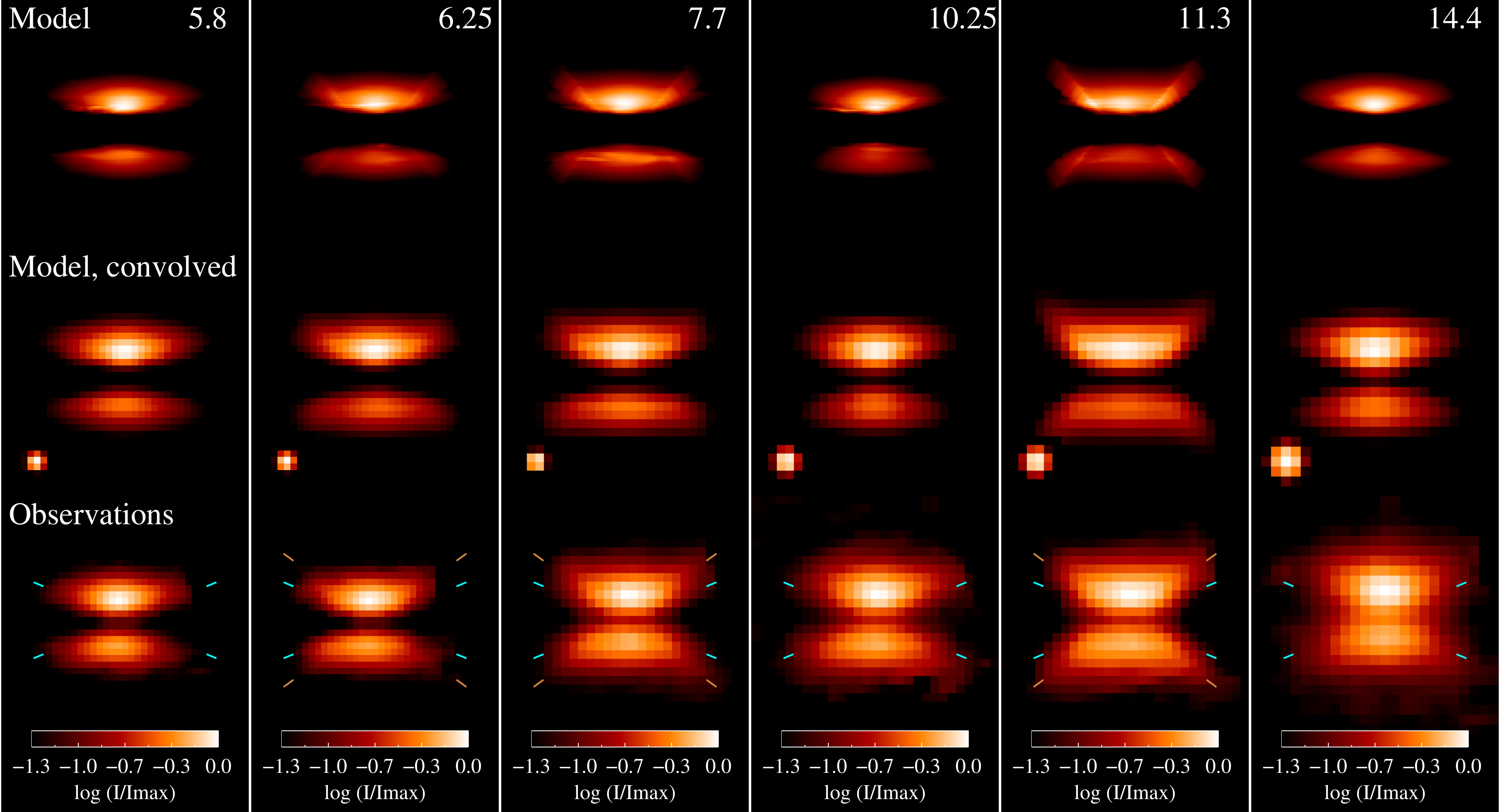}
\caption{Benchmark model including an astro-PAHs wind component. (upper) Model images at full resolution for selected wavelengths spanning the mid-infrared for bands including astro-PAHs contributions (6.25, 7.7, 11.3~$\mu$m) or bands representing adjacent bands with more continuum (5.8, 10.25, 14.4~$\mu$m) for comparison. (middle) Model images once convolved with JWST PSF. Note the alternance of more rounded shape for continuum dominated bands and chalice/hourglass shape for the one with astro-PAHs contribution. (lower row) observed MIRI images. The colorbars indicate the intensity levels, normalised
to the maximum intensity for each image, in log scale.}
\label{Figure_X_wings_transfer_appendix}
\end{center}
\end{figure*}
%
%
\begin{figure*}[!ht]
\begin{center}
\includegraphics[width=2\columnwidth,angle=0 , trim={0 0cm 0cm 0}, clip]{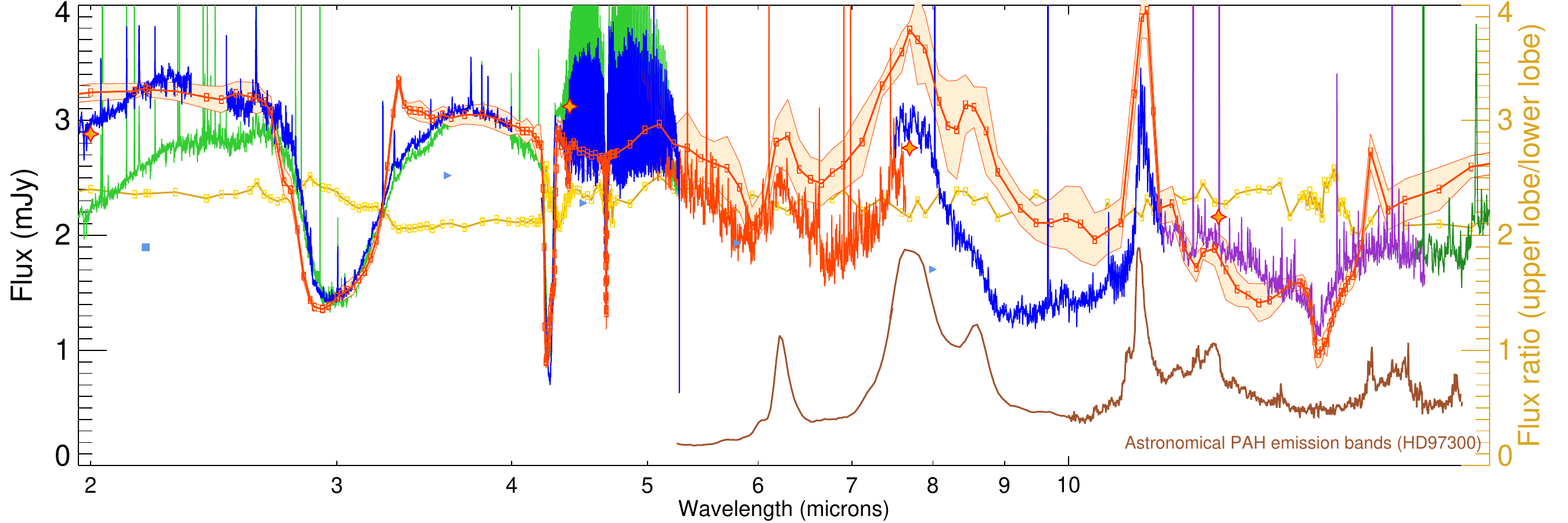}
\caption{Zoom of the benchmark model spectrum including the astro-PAHs grain model (orange) and estimated uncertainties on the calculation (light orange filled region) in the JWST spectral range, overplotted on the NIRSpec-MIRI combined spectrum. Note the emission in the model is higher in the 8-10 microns, which may be related to an overestimate of the silicates stretching mode contribution in emission in this range. The astronomical PAHs emission from the HD97300 disk is also shown in order to delineate the astronomical PAHs spectral profiles observed in another disk.}
\label{Figure_X_wings_transfer_model_spectrum}
\end{center}
\end{figure*}
%

\end{appendix}

\end{document}